\documentclass[aps,prx,reprint]{revtex4-1}

\usepackage{graphicx}
\usepackage{amssymb,amsmath}
\usepackage{epstopdf}
\usepackage{natbib}
\usepackage{array}
\usepackage{multirow}
\usepackage[shortlabels]{enumitem} 
\usepackage{color} 

\usepackage{tikz}
\usepackage{pgfplots}

\newcommand\nll{\mathrel{\mkern2.1mu\not\mkern-2.1mu\ll}}

\newcommand{\R}{\mathbb{R}}

\newcommand{\oneover}[1]{\frac{1}{#1}}
\newcommand{\eps}{\epsilon}
\newcommand{\bb}[1]{\mathbf{#1}}

\begin{document}

\title{
Free energy of singular sticky-sphere clusters
}
\author{Yoav Kallus }
\affiliation{Santa Fe Institute, 1399 Hyde Park Road, Santa Fe, NM 87501}
\author{Miranda Holmes-Cerfon}
\affiliation{Courant Institute of Mathematical Sciences, New York University, New York, NY 10012}

\begin{abstract}

Networks of particles connected by springs model many condensed-matter systems, from colloids interacting with a short-range potential, to complex fluids near jamming, to self-assembled lattices, to origami-inspired materials. Under small thermal fluctuations the vibrational entropy of a ground state  is given by the harmonic approximation if it has no zero-frequency vibrational modes, yet such singular modes are at the epicenter of many interesting behaviors in the systems above. We consider a system of $N$
spherical particles, and directly account for the singularities that arise in the sticky limit where the pairwise interaction is strong and short ranged.
Although the contribution to the partition function from singular clusters diverges in the limit, its asymptotic value can be calculated and depends on only two parameters, characterizing the depth and range of the potential. The result holds for systems that are second-order rigid, a geometric characterization that describes all known ground-state (rigid) sticky clusters.

To illustrate our theory we address the question of emergence: how does crystalline order arise in large systems when it is strongly disfavored in small ones?
We calculate the partition functions of all known rigid clusters up to $N\leq 19$, and show the
cluster landscape is dominated by hyperstatic clusters (those with more than $3N-6$ contacts) -- singular and isostatic clusters are far less frequent, despite their extra vibrational and configurational entropies. Since the most hyperstatic clusters are close to fragments of a close-packed lattice, this underlies the emergence of order in sticky-sphere systems, even those as small as $N=10$.

\end{abstract}


\maketitle

Particles that live on the mesoscale commonly interact over ranges much shorter than their diameters. 
Naturally-occurring systems include $C_{60}$ molecules \cite{Hagen1993, Doye2001} and colloids interacting via depletion \cite{Lu:2013dn,Manoharan:2015ko}. 
A widely studied synthetic system is colloidal particles coated with strands of DNA or other functional tethers, which create highly specific interactions with range roughly the length of a tether \cite{Nykypanchuk:2008cp,Dreyfus:2009gl,Rogers:2011et,Macfarlane:2011fh,DiMichele:2013bw}. 
Rather than modeling the details of the interparticle interaction, which are often not important for macroscale observations, 
it is convenient to model the interaction in the \emph{sticky limit}, which considers a central-force potential with a single well
that is narrow and deep \cite{Baxter1968,Holmes2013}.

\begin{figure}
\begin{tabular}{c c}
\includegraphics[width=0.45\linewidth]{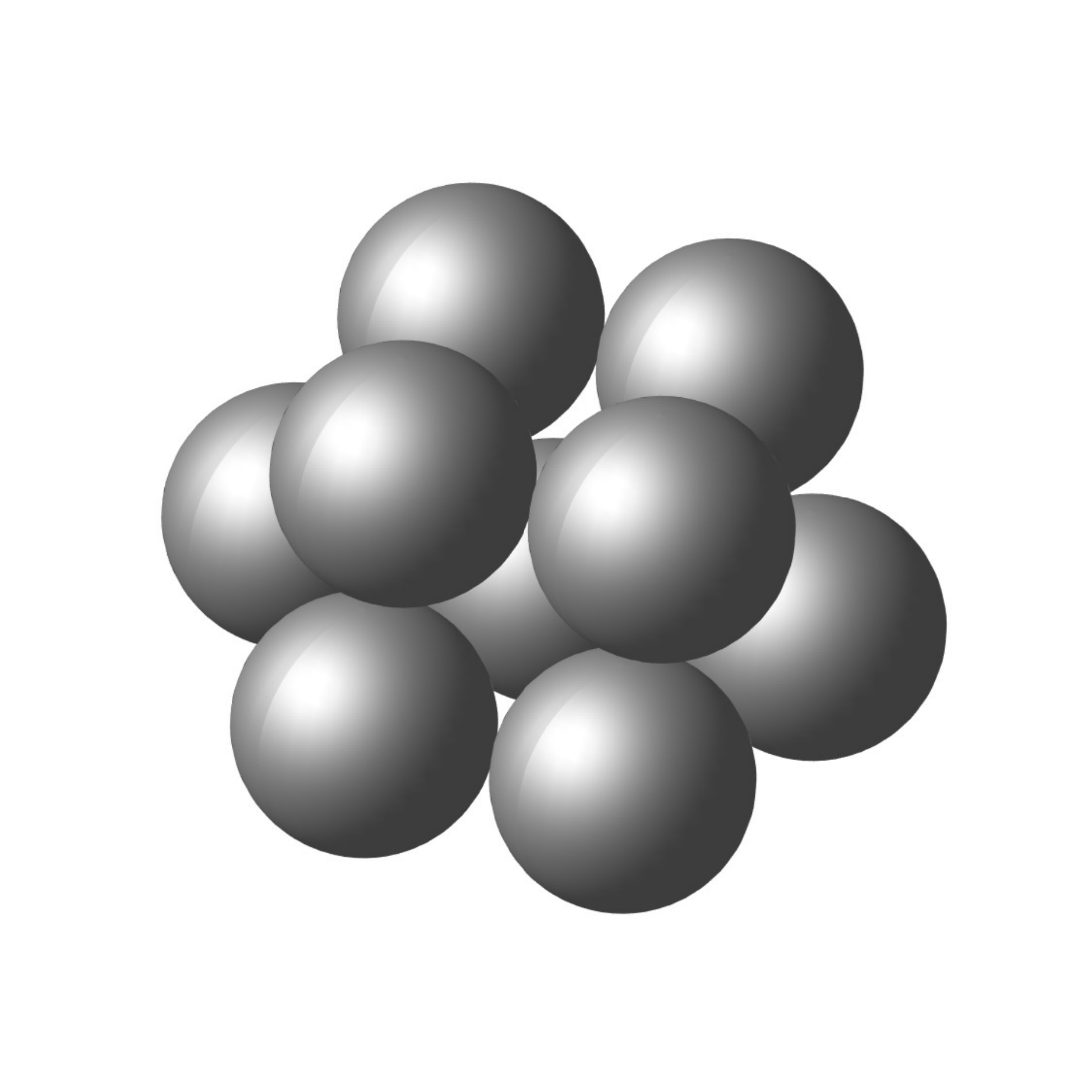} \hfill
\includegraphics[trim={3cm 3cm 3cm 3cm},clip,width=0.45\linewidth]{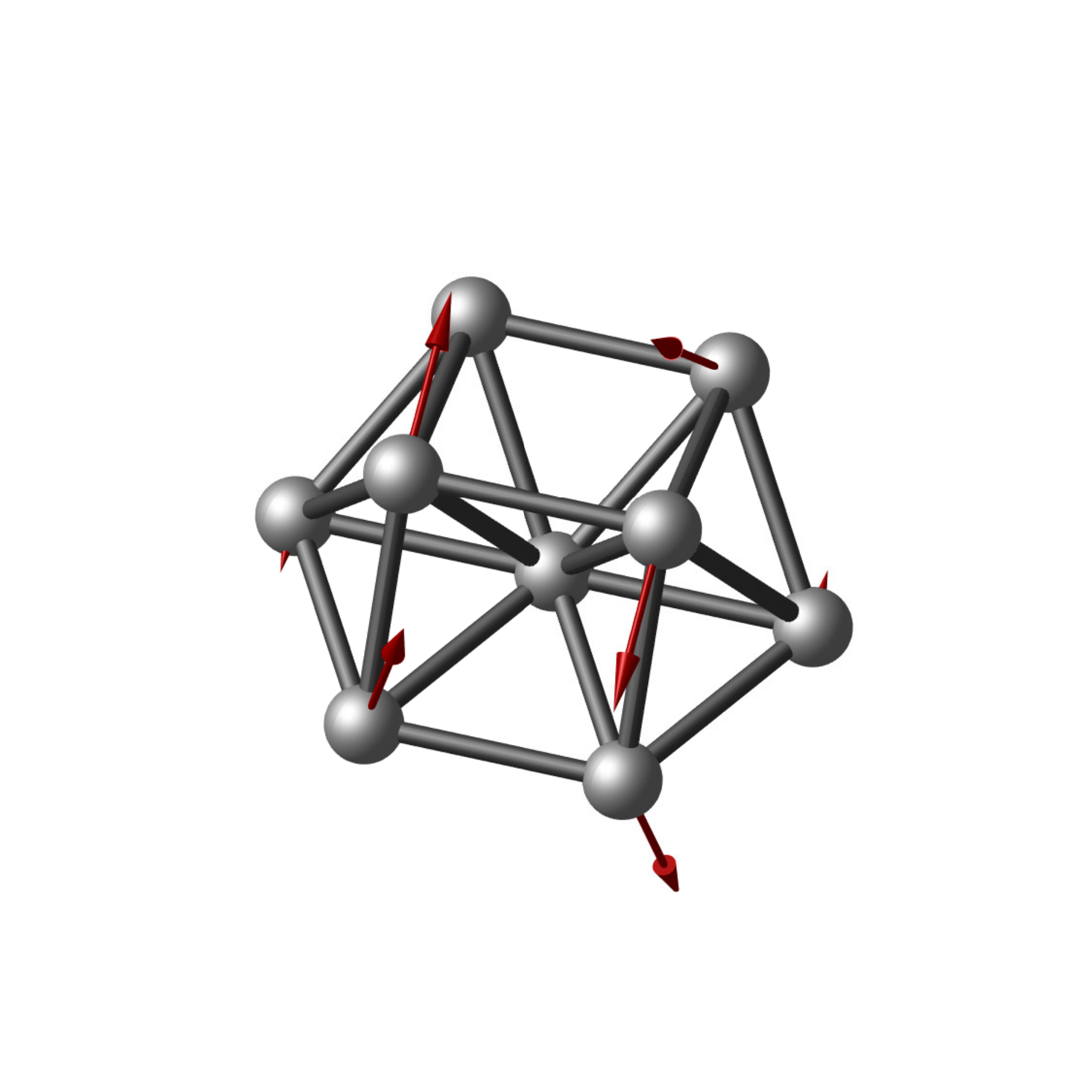} 
    \end{tabular}
\caption{\label{fig:n9}Left: the smallest singular cluster of sticky spheres, consisting of 9 spheres. Right: a framework representation
    of the cluster with the singular mode indicated by red arrows. The singular mode is a vibrational mode with zero frequency, which cannot be extended to any finite deformation without incurring an energy cost, or equivalently without breaking a  contact. }
\end{figure}

As the range of the interaction goes to zero, 
the space of energy-preserving motions available
to a finite system of particles is identical to that of a bar framework, with nodes located at the particle
centers and bars connecting pairs of particles that are touching, as in Figure \ref{fig:n9}.
The energy of the system is proportional to the number of contacts, or bars in the framework. When thermal fluctuations
allow the structure to slightly deviate from this space, the resulting entropy can be derived from the harmonic response of the network, provided all vibrational modes have
nonzero frequency.
The result contains a dimensionless geometric factor and a single physical parameter,
characterizing the temperature and stiffness of the interaction potential \cite{Holmes2013}.

This limit is appealing because it clearly separates the effects of geometry of the particles from those of the interaction potential \cite{Perry:2015ku,Perry:2016gk}, allowing one to more easily vary the latter to study self-assembly processes, for example. 
It was originally considered by Baxter to study phase transitions in fluids of particles with short-range interactions \cite{Baxter1968}, though singularities related to the ones we will solve for have prevented it from being more widely used \cite{Stell1991}.
The sticky limit can also be used as a controlled approximation for the entropy of a system of hard particles without attraction
near close packing \cite{Elser2014}. 

A natural starting point for investigating the consequences of the sticky limit 
is to consider a collection of finitely many spheres. In the sticky limit the lowest-energy states have the maximum number of contacts, and they are typically mechanically rigid. 
A recent body of work has focused on computing the set of rigid clusters of $N$ spheres and understanding their
thermodynamic properties \cite{Meng2010,Arkus:2011tl,Hoy:2012cr,Hoy:2015hz,HolmesCerfon:2016wa}. This program aims partly at identifying a set of geometrical possibilities for processes like
self-assembly \cite{Hormoz:2011ir,Zeravcic:2014it} and self-replication \cite{Zeravcic:2014ev}, and partly at understanding the question of emergence:
how does a finite system of particles transition
from a small size, where the preferred order might be incompatible with crystalline order, to a large one where it assembles into a highly structured
close-packed lattice \cite{Frank1952,Bernal1960,Doye1996,Manoharan2003,Hopkins2011,Teich2016}?

A major roadblock for such a program is the treatment of \emph{singular} clusters, those with vibrational modes with zero frequency  that do not extend to finite motions (singular modes), 
so-called because they correspond to singularities of a system of algebraic equations.
Current methods for treating the sticky limit do not work directly for these systems because the entropy associated with the singular modes diverges.
The smallest singular cluster occurs at a mere $N=9$ \cite{Arkus:2011tl}, preventing the sticky limit from being applied to systems that size or larger with current methods.
Yet singular clusters are not a rarity that can be glibly ignored: they account for about 2.5\% of the known rigid clusters of $N$ spheres \cite{HolmesCerfon:2016wa}.
A related problem is \emph{hyperstatic} rigid clusters, those with more than the $3N-6$ contacts required for a cluster to be rigid generically. The free energy of hyperstatic clusters also diverges in the traditional sticky limit, a fact which has  hindered the limit from being more widely applied to study bulk systems  \cite{Stell1991}.
The smallest hyperstatic cluster has $N=10$ spheres \cite{Arkus:2011tl}. Observations of colloids with a short-range depletion interaction found the $N=9$
singular cluster occurred with disproportionate frequency, and for $N\geq 10$ the majority of observed clusters were either singular or hyperstatic \cite{Meng2010}. 
This suggests there could be a competition between the higher entropy of a singular mode, and the lower energy of an additional contact. We wish to evaluate this competition and determine if singular clusters could be thermodynamically stable 
as the number of spheres increases into the bulk regime. 

In this paper we extend the sticky limit to systems with singular modes and non-isostatic numbers of contacts. We show that, provided the system is \emph{second-order rigid}, its partition function depends on only two parameters, characterizing the depth and range of the pair potential. This is only one more parameter than is required in the nonsingular case.
The property of being second-order rigid is explained in the text and characterizes all known rigid clusters.
Our approach is to calculate the partition function from an expansion of the potential energy function
up to fourth-order in the particle displacements. 
The result diverges as the range of the potential approaches zero, but by using the leading asymptotic term,
we obtain a finite result if the range is taken as a small, nonzero parameter.

This computation makes specific predictions for the probabilities of observing singular clusters in equilibrium, which we confirm by
comparing to simulations of particles interacting with a short but finite-range potential. 
We then calculate the free energies of all known rigid clusters of spheres up to $N\leq 14$ and of hyperstatic clusters for $15\leq N \leq 19$,
and show that the added entropy of singular clusters 
never beats the lower energy of the maximally hyperstatic clusters at realistic values of stickiness, except when the singular clusters are maximally hyperstatic.
Since the most hyperstatic clusters are fragments of a close-packed lattice (or close to such),
this observation explicitly and quantitatively demonstrates the emergence of order even at small system size. 
Finally, our results suggest a universal scaling, not yet explained, of the configurational and vibrational entropies of sticky sphere clusters
as a function of the number of contacts.


The problem of calculating the effects of singular modes extends far beyond the system specifically
under study in this paper.
Any system whose motions are limited by soft constraints, in the limit that the constraints become very stiff, could exhibit infinitesimally free motions that do not extend to finite degrees of freedom. 
In particular, singular modes exist and seem to play an important role in the mechanical and thermodynamic stability of many metamaterials,
a class of systems of growing interest in material
science, composed of mesoscopic or macroscopic building blocks and designed to have properties that are hard or impossible to engineer in traditional
materials. For example, metamaterials based on ball-and-spring networks may be designed to have singular modes on purpose so they fail under stress at
desired locations \cite{Paulose:2015hd}. 
Origami mechanisms designed to have a single extended degree of freedom may, as has been observed, end up with extensively many
infinitesimal degrees of freedom, with important implications to the mechanics of actuating the mechanism \cite{MuruganOrigami}.
As low-frequency modes seem to play an important role in the rheology of athermal fluids made of isotropic particles \cite{Gartner2016,Biroli2016},
zero-frequency modes, which occur in extensive number at the jamming point of anisotropic particles \cite{Donev2007}, could also have a significant effect.

While the mechanical effects above are athermal, zero-temperature effects, in this paper we focus on the thermal
effects of singular modes. Such effects are already relevant for some metamaterials, and will become relevant for others as the trend
toward miniaturization is followed and such systems are fabricated at smaller and smaller scales.
An abundance of low-frequency vibrational modes increases the entropy of a system, and can often be a crucial
factor stabilizing one structure over another when they are energetically equivalent: for example, such an effect
favors a face-centered-cubic close packing structure over a hexagonally-close-packed structure in many systems
\cite{Elser2014}, stabilizes marginally coordinated lattices in patchy colloids \cite{Mao2013}, and favors
an ordered structure over a zigzag structure in a rhombus lattice \cite{Mao2015}.
If given two energetically-equivalent structures, one has singular  modes,
this effect becomes even stronger: the entropic factor favoring the singular structure grows larger and larger
as the constraints become stiffer, as we show in this paper.
It is conceivable that in such cases the singular structure could be thermodynamically favored in an appropriate limit even when
it is energetically disfavored. We derive the conditions for such a scenario to occur, but 
we find that they are not present in the case of a system of $N$ sticky spheres for any $N\le19$.

\section{The sticky limit}

\subsection{Rigid clusters}

\begin{figure}
    \hspace{12mm}
    \scalebox{0.7}{
	\begin{tikzpicture}
	    \begin{axis}[
		    xlabel={$r$}, ylabel={$V(r)$},
		    xmin=0.5, xmax=2,
		    ymin=-10, ymax=10,
		    domain=0.5:2,
		    restrict y to domain = -10:30,
		    legend style={at={(0.97,0.97)},anchor=north east},
		]
		\addlegendimage{empty legend}
		\addplot[samples=400,dotted,color=black]plot (\x,{2.*exp((1-\x)/0.2)*(exp((1-\x)/0.2)-2)});
		\addplot[samples=400,dashed,color=black] {4.*exp((1-x)/0.1)*(exp((1-x)/0.1)-2)};
		\addplot[samples=400,color=black] {8.*exp((1-x)/0.02)*(exp((1-x)/0.02)-2)};
		\addlegendentry{$\mathcal E, \rho$};
	        \addlegendentry{$2,5$};
		\addlegendentry{$4,10$};
		\addlegendentry{$8,50$};
	    \end{axis}
	\end{tikzpicture}
    }
    \caption{\label{fig:morse}The sticky sphere limit of a Morse pair potential, $V(r) = \mathcal{E} e^{-\rho (r-1)}(e^{-\rho(r-1)}-2)$,
        where the depth $\mathcal E \to \infty$ and the range $\epsilon = 1/\rho \to 0$.}
\end{figure}
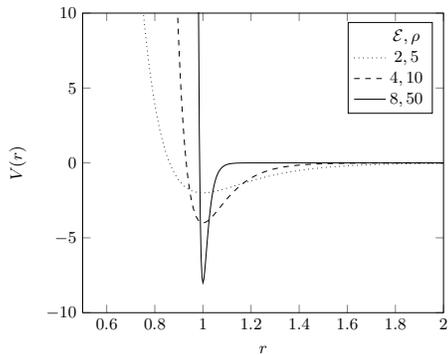

We start by explaining how to calculate the free energy in the sticky limit for a cluster of $N$ identical spheres with diameter $d$,
forming a cluster $\mathbf{r} = \bb{r}_1\oplus\bb{r}_2\oplus\ldots\oplus\bb{r}_N\in\R^{3N}$, where $\bb{r}_i\in\R^3$ is a
vector specifying the position of the center of the $i$th sphere, and $\mathbf{r}$ is the corresponding point in a $3N$-dimensional
configuration space. 
We suppose the cluster has $B$ \emph{bonds} between pairs of interacting spheres $E = \{(i_1,j_1),\ldots,(i_B,j_B)\}$. 
When the range of interaction is infinitesimally short, each bonded pair will be exactly touching, so the cluster lies in the solution set to the system of equations
\begin{equation}\label{eq:bonds}
    |\bb{r}_{i} - \bb{r}_{j}|^2 = d^2, \qquad (i,j)\in E\text.
\end{equation}
We focus for illustration on clusters that are rigid, though the calculations extend naturally to floppy ones. 
A cluster is defined to be \emph{rigid} if it lies on a connected component of the solution set to \eqref{eq:bonds} that contains only rotations and translations \cite{Asimow:1978en,secondrigidity}.
Physically, being rigid means one cannot continuously deform the cluster
internally by any finite amount while maintaining all contacts. 
This is a nonlinear notion that is \emph{not} equivalent to counting the number of infinitesimal degrees of freedom, as in the Maxwell-Calladine theorem
\cite{Calladine1978}. 

An example of a cluster that is rigid but has an infinitesimal degree of freedom is shown in Figure 
\ref{fig:n9}. This is a cluster of $N=9$ spheres, consisting of two bipyramids that share a sphere at a common vertex and are connected by three parallel bonds.
When one bipyramid is rotated along the axis going through the shared vertex and parallel to the
bonds,  the three bond lengths do not change to linear order in the displacement, so this rotation is an infinitesimal degree of freedom. The bond lengths do change at quadratic order in the displacement. In fact, any finite motion that is not a rigid body motion changes at least one bond length, and the cluster is rigid.

We suppose the potential energy of the system is
$U(\bb{r}) = \sum_{1\le i< j\le N}V(|\bb{r}_i-\bb{r}_j|)$, a sum of pair potentials $V(r)$ that depend on the distance $r$ between each pair. 
The pair potential is assumed to have a minimum at $d$ and approach zero rapidly as $r-d$ exceeds a certain characteristic interaction length $\eps$.
For $r-d\ll-\eps$, we assume the potential is positive and much larger in magnitude than its minimum value.
As a particular example, we consider the Morse potential $V(r) = \mathcal{E} e^{-\rho (r-1)}(e^{-\rho(r-1)}-2)$, where the pair potential has a minimum
at $d=1$, and the parameter $\rho$ determines the range of the interaction, $\epsilon = 1/\rho$.

The sticky limit occurs when the well of the pair potential is both \emph{narrow} and \emph{deep}. 
This limit can be constructed from any interaction potential with reasonable decay, by
simultaneously rescaling the lengthscale $r-d$ and the depth of $V(r)$ (see Figure \ref{fig:morse}).
The sticky limit has traditionally been considered under scalings such that the contribution of the well to the
partition function for a single pair of spheres approaches a constant \cite{Baxter1968,Stell1991}.
Using Laplace's method (see SI section \ref{sec:calcs}) this constant is shown to be proportional to 
\begin{equation}\label{eq:kappa}
\kappa = \lim_{\eps\to0} \frac{\sqrt{2\pi}}{d}\frac{e^{\beta \mathcal E}}{(\beta a)^{1/2}}
\end{equation}
where $\mathcal E = -V(d)$, $a=V''(d)$, 
and $\beta=(k_BT)^{-1}$ is the inverse of temperature $T$ times the Boltzmann constant.
Since it represents the equilibrium tendency of of spheres to stick
to each other, it is called  the \emph{sticky parameter} \cite{Holmes2013}. 
s
The sticky parameter characterizes the thermodynamic trade off between the free energy of a particle in a bulk fluid of fixed
packing fraction $\phi$, $\exp (-\beta F_\text{fluid} / N) \sim d^3/\phi$, and a particle in an isostatic network of bonded particles,
$\exp (-\beta F_\text{iso}/N) \sim (\beta a)^{-3/2} e^{3 \beta \mathcal E}$. Therefore, the limit $\eps\to0$ as $0<\kappa<\infty$
is relevant for studying the instability of a fluid toward forming an isostatic bonded network \cite{Charbonneau2007}. However, a close-packed
network has six bonds per particle in the bulk limit and so the relevant limit for studying its formation is
$0<\lim_{\eps\to0} (\beta \alpha)^{-1/2} e^{2\beta\mathcal E}<\infty$, which requires $\kappa\to0$. As the number of spheres $N$ in a finite system rises,
the relevant clusters might vary from isostatic to close-packed, so one might wish to consider different limits.
Our calculation allows the limits $\eps\to0$ and $\mathcal E\to\infty$
to be taken in any way desired, as we show that it does not change the form of the leading term in the partition function.

The contribution to the canonical partition function from each rigid cluster is given by
\begin{equation}\label{eq:Ze}
    Z_{\mathbf{r}} = \int_{\mathcal N_{\mathbf{r}}} \exp[-\beta U(\mathbf{r}')] d^{3N}\mathbf{r}'\text,
\end{equation}
where $\mathcal N_{\mathbf{r}}$ is an integration domain
constructed by taking all configurations corresponding to the cluster of interest, including translations,
rotations, permutations, and reflections, and fattening it by a length that is small compared to the particle diameter $d$
and large compared to the interaction range $\epsilon$ (see SI section \ref{sec:calcs}.)
We take the total partition function to be the sum of the contributions from all the rigid clusters, ignoring floppy clusters in this paper.
The free energy is $F_\mathbf{r} = -\beta^{-1}\log Z_\mathbf{r}$. 

We may remove rigid-body degrees of freedom by a change of variables. 
If we let $W\subset \R^{3N}$ 
be the linear subspace of infinitesimal rigid-body motions, 
then we may restrict the integral to its orthogonal complement $W^\perp$, and include a factor $I$ equal to the square root of the determinant of the moment of inertia tensor 
 (SI section \ref{sec:calcs}). 
Summing over permutations contributes a factor of $N!/\sigma$, where $\sigma$ is the size of
the Euclidean symmetry group of the cluster \cite{Cates:2015ik}.
If the spheres lie in a container of volume $V$, then the free volume for each cluster is nearly $V$ as long as $V\gg d^3$. 

To evaluate the remaining parts of the integral in \eqref{eq:Ze} we expand each term $V(\bb{r}_{ij} + \delta \bb{r}_{ij})$ in powers of $\delta \bb{r}_{ij}$,
where we write $(\cdot)_{ij}$ for $(\cdot)_{i}-(\cdot)_{j}$.
The expansion up to second order in the displacements yields (see SI \eqref{eq:SIexpand} for detailed expansions and error bounds)
\begin{equation}\label{eq:U2}
    U(\mathbf{r}+\delta\mathbf{r}) = -B\mathcal{E} + \tfrac12\langle\delta\mathbf{r},M \delta\mathbf{r}\rangle \text,
\end{equation}
where $M$ is a symmetric linear map, whose matrix representation is usually referred to as the \emph{dynamical matrix}.
When \eqref{eq:U2} is plugged into the integral in \eqref{eq:Ze}, the integral converges if and only if $M$,
considered as a map $W^\perp\to W^\perp$, is positive definite, that is, all its eigenvalues are positive. The result
gives the harmonic approximation for the vibrational partition function. 

To see what the positive definiteness of the dynamical matrix implies about the rigidity of the cluster, let us consider
how one might deform a cluster along a path $\bb{p}(t)$ with $\bb{p}(0) = \bb{r}$ without changing any of the bond lengths.
Taking one derivative of \eqref{eq:bonds} shows that 
\begin{equation}\label{eq:R}
    (\mathbf{r}_{i}-\mathbf{r}_j)\cdot(\mathbf{p}'_{i}(0)-\mathbf{p}'_j(0)) = 0 \text{ for all }(i,j)\in E\text.
\end{equation} 
Defining the linear map $R:\R^{3N}\to\R^{B}$ that maps $\mathbf{p}'(0)$ to the $B$-dimensional vector with
entries given by the left hand sides of \eqref{eq:R}, we see that $\bb{p}(t)$ maintains the constraints \eqref{eq:bonds}
only if $\mathbf{p}'(0)$ is in the null space of $R$. The matrix representation of $R$ is known as the \emph{rigidity matrix} 
in the mathematical study of framework rigidity. An element of the null space of $R$ 
is called a \emph{flex}, and it is \emph{trivial} if it is a rigid-body motion (that is, in $W$). If the only flexes are trivial, then the cluster is said to be \emph{infinitesimally rigid} 
or \emph{first-order rigid}. This is sufficient for the cluster to be rigid in the nonlinear sense \cite{secondrigidity}. 
The dynamical map (matrix) and the rigidity map (matrix) are related as $R^*R = Md^2 /a$,
where $R^*$ is the adjoint (transpose), so the harmonic approximation converges exactly when the cluster is first-order rigid. 

The harmonic approximation fails when the null space of the dynamical matrix extends beyond $W$,
or equivalently when nontrivial flexes of the rigidity matrix exist. These are infinitesimal degrees of freedom that may or may not be
extendable to finite degrees of freedom. If they are not, we call them \emph{singular} directions (or modes or flexes.) The cluster in Figure \ref{fig:n9}
has a singular direction corresponding the twisting motion described above.

To integrate along these singular directions we continue the expansion of the potential energy function.
Let $X$ be the space of singular direction: $X=\mathrm{ns} (M)\cap W^\perp$, where $\mathrm{ns}(\cdot)$ denotes the null space. 
Let $Y=(\mathrm{ns} (M))^\perp$ be its orthogonal complement in $W^\perp$. 
For every $\delta \bb{r} \in W^\perp$ we may write $\delta \bb{r} = \bb{x} + \bb{y}$ where $\bb{x} \in X$, $\bb{y}\in Y$.
We plug this decomposition into the potential energy function, use the fact that $\bb{x}_{ij} \cdot \bb{r}_{ij} = 0$ for all $(i,j)\in E$ by \eqref{eq:R},
and keep terms up to $O((|\mathbf x|^2+|\mathbf y|)^2)$ (see SI \eqref{eq:vij-xy} for these calculations and error bounds). 
We obtain, neglecting the error terms,
\begin{multline}
    U(\mathbf{x},\mathbf{y}) = 
    -B\mathcal E \\ + ad^2 \tilde{U}_0(\mathbf{x}) +
                                      ad\langle\tilde{\mathbf{u}}_1(\mathbf{x}),\mathbf{y}\rangle + 
				      \tfrac12 a \langle\mathbf{y},\tilde{M} \mathbf{y})\text.
\end{multline}
Here $\tilde M = M/a$, $\tilde U_0(\bb{x})$ is a real scalar, and $\tilde{\bb{u}}_1(\bb{x})\in Y$ is a $3N$-dimensional vector, constructed as 
\begin{align}
    \tilde U_0(\bb{x}) &= \sum_{(i,j)\in E} \frac{1}{8d^4} (\mathbf{x}_{ij}\cdot\mathbf{x}_{ij})^2 \\
    [\tilde {\mathbf{u}}_1(\bb{x})]_i &= \sum_{j\text{ s.t. }(i,j)\in E} \frac{1}{2d^3} (\mathbf{x}_{ij}\cdot\mathbf{x}_{ij}) \mathbf{r}_{ij} 
\end{align}
If $U(\bb{x},\bb{y})-U(0,0)$ is positive for all nonzero $(\bb{x},\bb{y})$  we can calculate
the leading-order term of the partition function (see SI \eqref{eq:SIintx}):
\begin{multline}\label{eq:intx}
	Z_{\mathbf r}=\\
	\frac{ I V e^{\beta B\mathcal{E}}}{(\det \tilde{M}|_Y)^{1/2}\sigma} \left(\frac{2\pi}{a\beta}\right)^{\frac{d_Y}{2}}
	\left(\frac{d^2}{a\beta}\right)^{\frac{d_X}{4}} \left(\int_{X}e^{-Q(\tilde{\mathbf{x}})}d\tilde{\mathbf{x}}\right)
	\text,
\end{multline}
where $d_X, d_Y$ are the dimensions of $X$ and $Y$, and
\begin{align}\label{eq:Q}
Q(\tilde{\mathbf{x}}) &= 
\frac{1}{ad^2} \min_{\mathbf{y}\in Y} U(d\tilde{\mathbf{x}},\mathbf{y}) \nonumber\\
&= 
\tilde{U}_0(d\tilde{\mathbf{x}}) - \tfrac12\langle\tilde{\mathbf{u}}_1(d\tilde{\mathbf{x}}),\tilde{M}^{-1}\tilde{\mathbf{u}}_1(d\tilde{\mathbf{x}})\rangle
\end{align}
is a dimensionally reduced quartic form that gives the minimum of the potential at a fixed displacement along
the singular directions.
Letting $\gamma=e^{\mathcal \beta \mathcal E}$,  $\alpha = (a\beta d^2)^{1/4}$, and $B_{\text{ISO}} = 3N-6$, we can write \eqref{eq:intx} as 
\begin{equation}\label{eq:Zfinal}
Z_{\mathbf r} = Vd^{3N-3}(2\pi)^{-\frac{B_\text{ISO}}{2}} \gamma^{B} \alpha^{-2B_\text{ISO}+d_X} z_{\mathbf r} , 
\end{equation}
where
\begin{equation}\label{eq:zg}
    z_{\mathbf r} = \frac{I}{\sigma d^3}(\det \tilde{M}|_Y)^{-\tfrac12} (2\pi)^{-\frac{d_X}{2}} \int_{X}e^{-Q(\tilde{\mathbf{x}})}d\tilde{\mathbf{x}}.
\end{equation}

Equations (\ref{eq:Zfinal}--\ref{eq:zg}) are our main result. They express the partition function for a sticky rigid cluster as a product of a number of dimensional quantities which
depend on the pair potential, temperature, and particle dimensions, times a dimensionless quantity $z_{\mathbf r}$. We call $z_{\mathbf r}$ the \emph{geometrical partition function} because it depends
only on the positions of the particles, which are given by solving the geometry problem defined by equation \eqref{eq:bonds}, and not on any externally controllable parameters. 
The interaction potential and temperature enter only in combination through the two parameters $\gamma$ and $\alpha$, which measure, respectively, the contribution of each extra bond and each singular direction. 
Remarkably, only the second derivative of the interaction potential affects the free energy. All higher-order derivatives contribute to subleading order. 

When does the remaining integral in \eqref{eq:zg} converge? We will show it converges for all clusters that
are \emph{second-order rigid}. To explain what this means, consider again the hypothetical deformation of the cluster $\bb{p}(t)$,
and now take the second time-derivative of \eqref{eq:bonds}:  
\begin{equation}\label{eq:Rp}
    (\bb{r}_i - \bb{r}_j)\cdot(\bb{p}''_i(0)-\bb{p}''_j(0)) + \|\bb{p}'_i(0)-\bb{p}'_j(0)\|^2 = 0 
\end{equation}
for all $(i,j)\in E$.
A given flex $\mathbf{p}'(0)$ can be extended to a second-order
motion only if there is a solution $\mathbf{p}''(0)$ to the linear equation \eqref{eq:Rp}.
If such a solution exists, the pair $(\mathbf{p}'(0), \mathbf{p}''(0))$ is called a second-order flex.
If there is no nontrivial second-order flex, then the cluster is second-order rigid.
Second-order rigidity is a sufficient condition for rigidity \cite{secondrigidity}.
However, analogous higher-order versions do not necessarily imply rigidity \cite{higherrigidity}.

Since  $U(\mathbf{x},\mathbf{y}) - U(0,0) = \sum_{(i,j)\in E} \frac{a}{2d^2}[ (\mathbf{r}_{ij}\cdot\mathbf{y}_{ij})
+ \tfrac12 (\mathbf{x}_{ij}\cdot\mathbf{x}_{ij})]^2$, then 
$U(\mathbf{x},\mathbf{y}) = U(0,0)$ for some $\mathbf{x}\in X$ and $\mathbf{y}\in Y$ if and only if
$\mathbf{r}_{ij}\cdot\mathbf{y}_{ij} + \tfrac12 (\mathbf{x}_{ij}\cdot\mathbf{x}_{ij})=0$
for all $(i,j)\in E$, namely, $(\mathbf{x},2\mathbf{y})$ is a second-oder flex. If the cluster is second-order rigid then no such flex exists,
so $U(\mathbf x,\mathbf y)>U(0,0)$ for all nonzero $(\mathbf x,\mathbf y)$, and \eqref{eq:zg} converges.

\subsection{Square well potential}
We have assumed a potential with nonzero second derivative at the minimum, but  
Baxter's original sticky-sphere limit considered a square well potential of depth $\mathcal E$
and width $\eps$, where 
the derivatives $V^{(n)}(d)$ vanish for all $n$ \cite{Baxter1968}. However, we can show (SI section \ref{sec:squarewell}) that while this difference
changes the prefactors in the calculation, it does not change how the partition function scales with the parameters $\gamma$ and $\alpha$,
which are defined for a square well potential to be $\gamma=e^{\beta\mathcal E}$, $\alpha=(d/\eps)^{1/2}$. 
Therefore, our calculations are a natural extension of Baxter's limit. 

\subsection{Floppy clusters}

We have derived \eqref{eq:Zfinal} for rigid clusters, but a similar calculation can be performed for floppy clusters, those with internal degrees of freedom along which the cluster can deform by some finite amount. In this case, some of the zero vibrational modes extend to finite degrees of freedom, and some do not so they are singular directions. 
We suppose the internal degrees of freedom form a manifold $\mathcal M$ with dimension $m$, and and that the number of singular modes $d_X$ is constant over the manifold. 
Now $X$, $Y$ are the linear subspaces along which the cluster is rigid with $X$ representing the singular directions and $Y$ the non-zero vibrational modes. 
The zero vibrational modes are the union of $X$ and the tangent space to $\mathcal M$. 
If $\mathcal M$ has dimension $m$, then we must have 
 $\text{dim}(X) + \text{dim}(Y) + m = 3n-6$. 

The additional calculations to deal with the floppy degrees of freedom closely resemble those performed in \cite{Holmes2013} and result in an integral over the internal degrees of freedom of a cluster. The resulting partition function still has the form \eqref{eq:Zfinal}, where the geometric partition function is now $z_{\mathbf r}^{\text{tot}}$ given by 
\begin{equation}\label{eq:floppy}
z_{\mathbf r}^{\text{tot}} = \int_{\mathcal M} z_{\mathbf r}(\mathbf{x})d\sigma_{\mathcal M}(\mathbf{x}),
\end{equation}
with $z_{\mathbf r}(\mathbf{x})$ given by \eqref{eq:zg}, and in both \eqref{eq:Zfinal}, \eqref{eq:zg} we must set  $B_{ISO}=3N-6-m$. 
Here $d\sigma_{\mathcal M}$ is the natural surface measure on the manifold, described in detail in \cite{Holmes2013}. 

\section{Free energy landscape of small clusters}

Equation \eqref{eq:Zfinal} shows that extra bonds and singular directions contribute to the stability of a rigid cluster by factors, $\gamma$
and $\alpha$, that both diverge in the sticky limit. This observation raises some natural questions:
for finite values of these parameters, which kinds of cluster tend to dominate the partition function? 
How does this answer depend on $N$? And, with a sight toward emergence, how do close-packing fragments come to dominate the landscape as $N\to\infty$? 

We address these questions by calculating the geometrical partition functions for all known rigid clusters (see SI section \ref{sec:numerical} for methods). 
We use the lists of rigid clusters produced by Holmes-Cerfon \cite{HolmesCerfon:2016wa}, which enumerated rigid clusters up to $N=14$, and rigid clusters containing a given minimum number of contacts up to $N=19$.
These are thought to be nearly complete lists of rigid clusters for each $N$ and for each specified number of contacts. All clusters in the lists are second-order rigid to numerical tolerance, so \eqref{eq:Zfinal} converges. 
We use the results to characterize the competition between singular and hyperstatic clusters at each value of $N$.

\subsection{$N\leq 8$} All clusters of these sizes are regular (isostatic and nonsingular). 
For $N\le5$, there is only a single rigid cluster for each value of $N$: a single sphere,
a dimer, a triangle, a tetrahedron, and a triangular bipyramid. For $N=6,7,$ and $8$, multiple rigid
clusters exist, all nonsingular, and all with the same number of bonds. 
The most important factor in the partition function is the symmetry number, with low-symmetry clusters dominating high-symmetry clusters 
\cite{Meng2010,Wales:2010jp,Calvo:2012bw}.

\subsection{$N=9$}
The smallest singular rigid cluster, illustrated in  Figure \ref{fig:n9}, occurs at $N=9$.
Including this cluster, there are $52$ rigid clusters, not counting enantiomers, all
with the same number of bonds \cite{Arkus:2011tl}.
So, the partition function 
is given, up to constant factors, by $Z = \alpha z_1 + \sum_{i=2}^{52} z_i$,
where $z_i$ is the geometric term in \eqref{eq:Zfinal} for the $i$th cluster, the singular
cluster being first. The equilibrium probability of the singular cluster is then 
\begin{equation}\label{eq:P1}
P_1 = \frac{\alpha}{K+\alpha}
\end{equation}
with $K=\frac{1}{z_1}\sum_{i=2}^{52} z_i\approx 235$.
In the sticky sphere limit, $\alpha\to\infty$, and this probability approaches
one. However, we expect this estimate to hold even for finite $\alpha$.

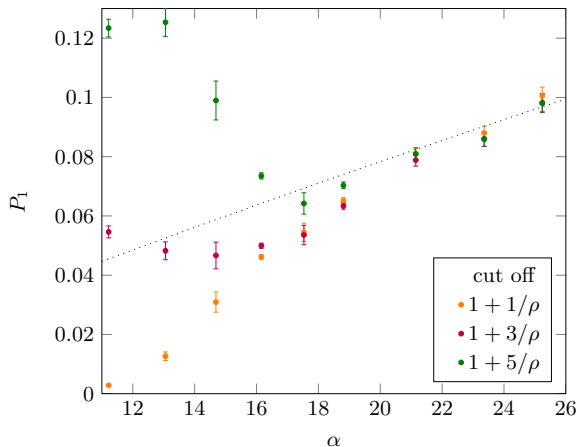
\begin{figure}
    \begin{center}
    	\scalebox{0.9}{\begin{tikzpicture}
	    \begin{axis}[
		    xlabel={$\alpha$}, ylabel={$P_1$},
		    xmin=11, xmax=26,
		    ymin=0, ymax=0.13,
		    domain=11:26,
		    legend style={at={(0.97,0.03)},anchor=south east},
		    legend entries={cut off,$1+1/\rho$,$1+3/\rho$,$1+5/\rho$},
		    yticklabel style = { /pgf/number format/fixed },
		]
		\addlegendimage{empty legend}
	        \addplot[only marks, mark size=1.0pt, error bars/y dir=both,error bars/y explicit,color=orange] table[x index=0, y index=1, y error index =2] {n9-k200.dat};
	        \addplot[only marks, mark size=1.0pt, error bars/y dir=both,error bars/y explicit,color=purple] table[x index=0, y index=3, y error index =4] {n9-k200.dat};
	        \addplot[only marks, mark size=1.0pt, error bars/y dir=both,error bars/y explicit,color=green!50!black] table[x index=0, y index=5, y error index =6] {n9-k200.dat};
		\addplot[dotted,color=black]plot (\x,{\x/(235.4+\x)});
	    \end{axis}
	\end{tikzpicture}}
\caption{Simulation results for $N=9$ sticky spheres with $\kappa\approx220$ and $\rho=30,40,50,60,70,80,100,120,$ and $140$ interpreted
    using different bond cut offs. The dashed line shows the theoretically predicted frequency of the singular cluster, \eqref{eq:P1}. \label{fig:sims}}
\end{center}
\end{figure}


To test this prediction, we sampled from the canonical ensemble of 9 spheres interacting via a Morse-harmonic potential with various ranges and depths using
a Monte Carlo simulation (see SI Section \ref{sec:sims} for methods). 
The diameter is taken to be $1$, and the range of the interaction is characterized by a parameter $\rho$.
For a given value of the range parameter $\rho$ the pair potential is active over a distance of about $5/\rho$, after which it has decayed to about 1.3\% of its minimum depth.
For each sampled structure we construct an adjacency matrix by specifying a cutoff distance for bonds
and identify the rigid cluster, if any, whose adjacency matrix is isomorphic to the constructed one. 
For small values of the range parameter $\rho$ (corresponding to a large interaction range),
this method was extremely sensitive to the cutoff being used, but for $\rho \gtrsim 70$ (corresponding to $\alpha \gtrsim 18$),
the calculated cluster frequencies converged and agreed with the theoretical prediction, as illustrated in  Figure \ref{fig:sims}.
 
The discrepancy for small $\rho$ likely arises because of interaction between non-nearest neighbors.
The rigid cluster geometries are minima of $U(\mathbf{r})$ only in the limit $\rho^{-1}\to 0$, and are deformed
for $\rho^{-1}>0$ due to interactions between spheres that are not in contact in the ideal structure.
The minimum gap in the rigid clusters of 9 spheres is $h_\text{min}\approx0.05d$, which occurs in seven of the 52 clusters, and 30 clusters
have gaps of $0.09d$ or less. When $5\rho^{-1}/h_\text{min}\nll 1$, the deformations of the local minima are substantial.
Small gaps also lead to problems in the identification of cluster geometries in our numerical sample, since they
could be identified as bonds.

Even though \eqref{eq:P1} applied only in the limit that $\alpha\to\infty$, and so $P_1\to1$, we observe remarkable agreement even when $P_1$
is only around 8\%.

Experiments that observed colloids interacting via a depletion interaction found the singular cluster occurred with frequency 11\% ( 95\% confidence interval 4\%-27\%) \cite{Meng2010}. A calculation neglecting the coupling between the
zero and nonzero modes predicted only a 3\% probability for the singular cluster \cite{Meng2010}, and the excess stability was surmised to correspond to the free energy of about half of an extra bond. 
The range of attraction was estimated to be about 1.05 times the particle diameter, which corresponds to a range parameter of $\rho\approx 30$ \cite{Malins:2009dt,Wales:2010jp}. The depth of interaction was estimated to be about $4k_bT$, but this gives a sticky parameter of $\kappa=1.6$ for which clusters should melt; a depth of $\approx 8 k_bT$ is probably closer to the truth so that $\kappa\approx 60$, and is consistent with recent simulations \cite{Fackovec:2016ds}. 
These values give $\alpha\approx11$ so the theoretically predicted probability is $P_1 \approx 0.045$, within the experimental confidence interval.
These parameters are in the regime where non-nearest neighbor interactions are relevant, so we cannot hope the calculation will agree exactly with the
experimentally observed value, but it is still notable that the two are consistent.

\begin{figure}
\begin{center}
\includegraphics[trim={0cm 0 1.5cm 0},clip,width=0.75\linewidth]{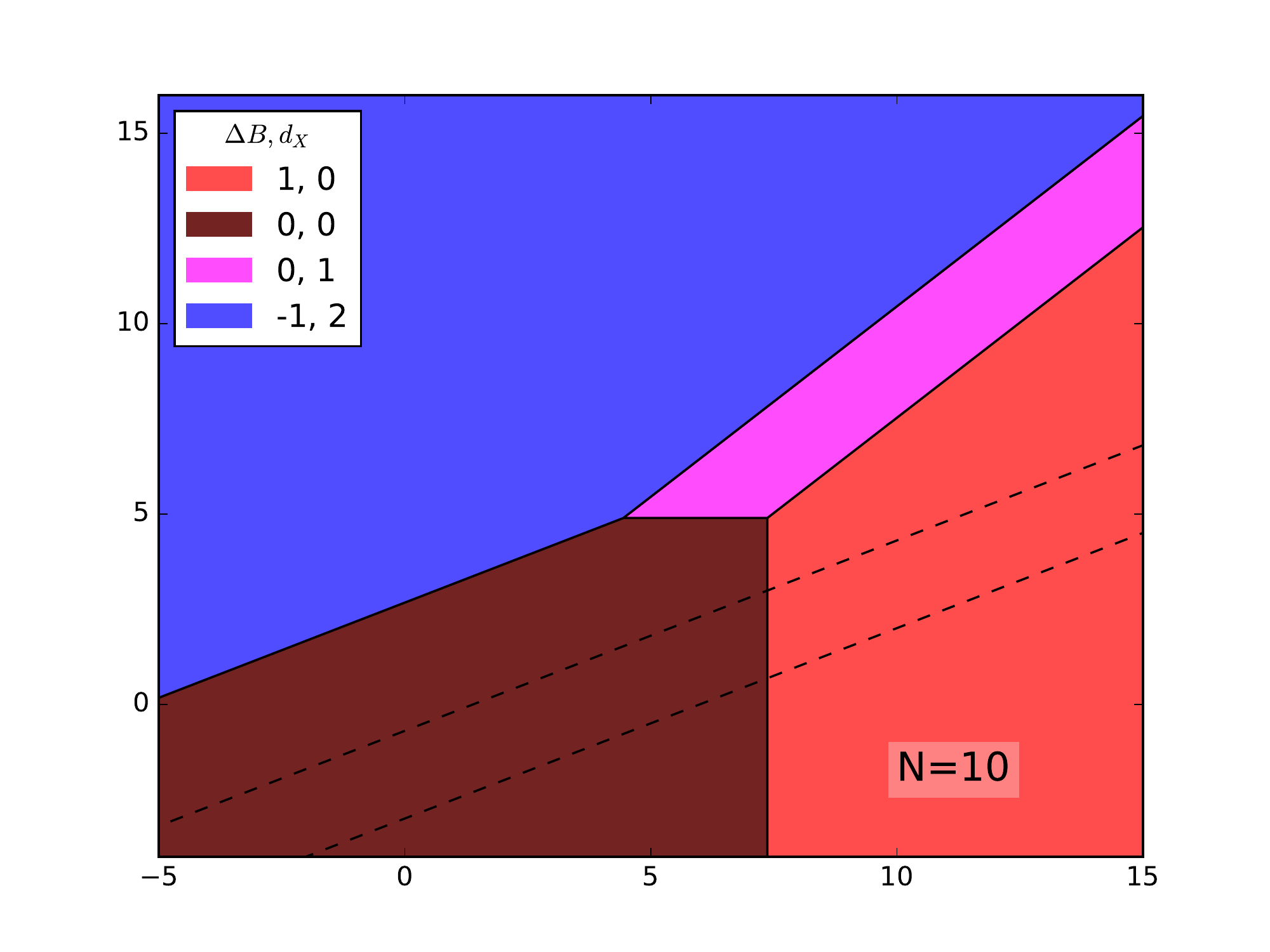}
\includegraphics[width=0.75\linewidth]{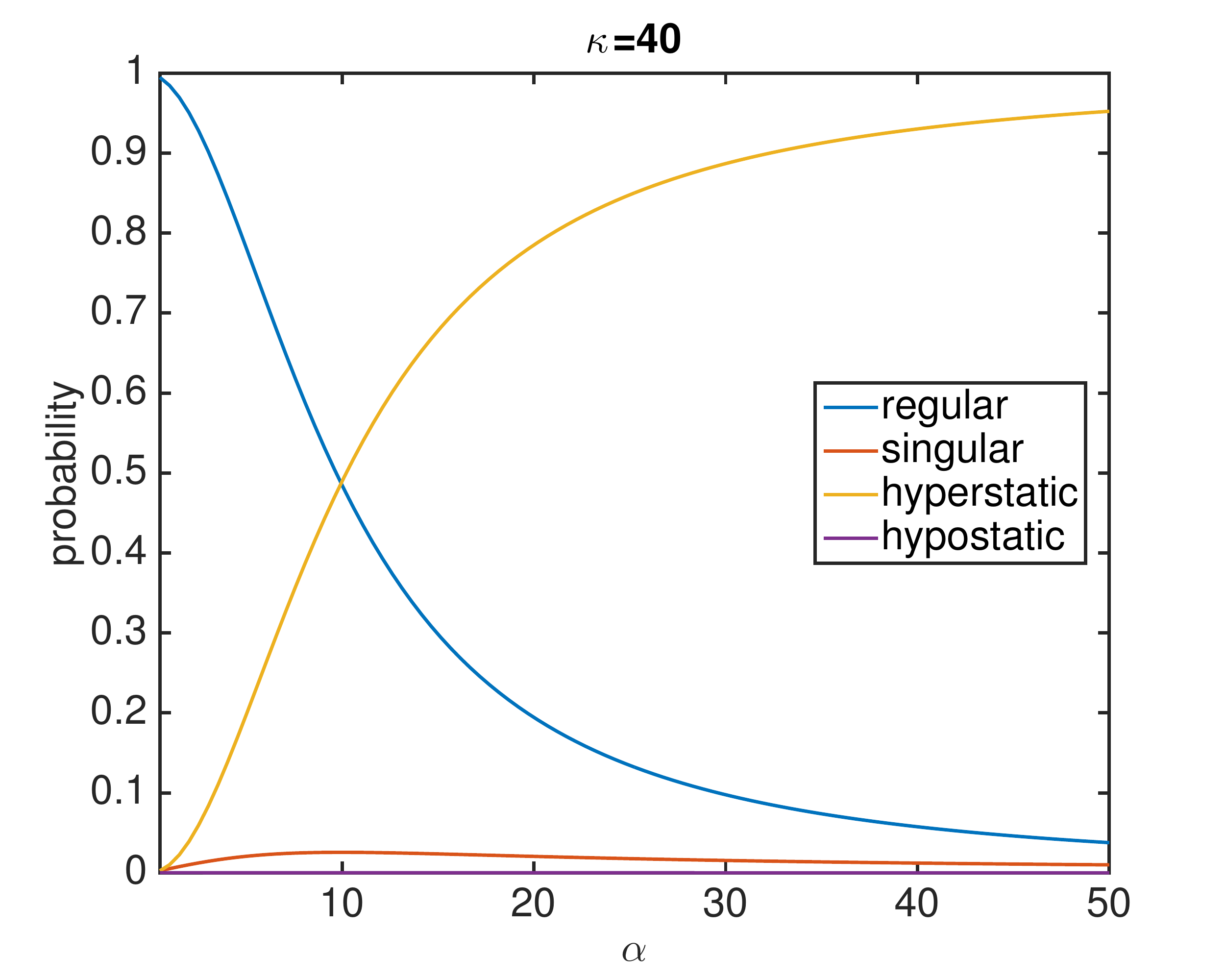}
\end{center}
\caption{Top: phase diagram for $N=10$ spheres. 
Colors indicate which kind of cluster has the largest partition function for each value of $\gamma$, $\alpha$, where the clusters are grouped by the number of extra contacts $\Delta B$ and the number of singular modes $d_X$. Dashed lines are $\kappa=10$ (top), $\kappa=1000$ (bottom). 
Bottom: Relative frequency of the types of rigid clusters of $10$ spheres as a function of $\alpha$ (roughly the inverse square root width of the potential), at $\kappa=40$.
\label{fig:region1}}
\end{figure}

\begin{figure}
\begin{center}
\includegraphics[trim={0cm 0 0 0},clip,width=\linewidth]{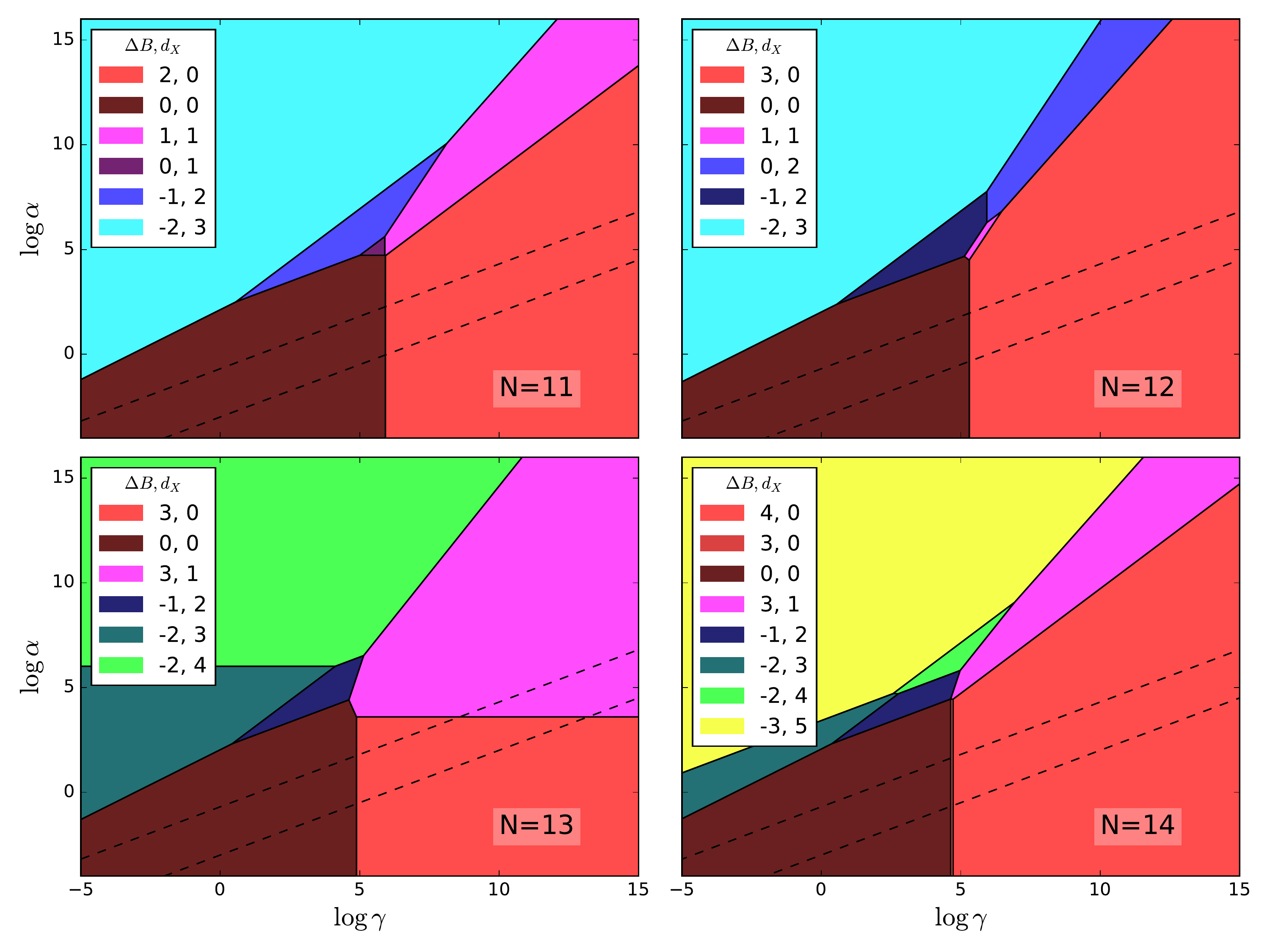}
\end{center}
\caption{Phase diagrams for rigid clusters of $N=11$--$14$ spheres. Labeling is the same as in  Figure \ref{fig:region1}. 
\label{fig:region2}}
\end{figure}

\subsection{$10\leq N\leq 19$} 
To analyze the landscape for higher $N$, we partition the rigid clusters based on the number of extra bonds $\Delta B=B-(3N-6)$
above or below the isostatic number, and the number of singular zero modes $d_X$. 
The canonical partition function is then
\begin{equation}\label{eq:Zpart}
    Z= \kappa^{3N-6}\sum_{\Delta B, d_X} \alpha^{d_X} \gamma^{\Delta B}z_{\Delta B,d_X} \text,
\end{equation}
where $z_{\Delta B,d_X}$ is the sum of the geometric contributions of all the clusters in a given partition (see SI Section \ref{sec:values} for values), and we have used the fact that the sticky parameter (\eqref{eq:kappa}) is
\begin{equation}
\kappa = \frac{\sqrt{2\pi}\gamma}{\alpha^2}.
\end{equation}
We analyze the landscapes for $N\geq 10$ by determining which term in \eqref{eq:Zpart} is the largest for different values of
$\alpha$ and $\gamma$.  

The phase diagram for $N=10$, shown in  Figure \ref{fig:region1}, indicates where on the parameter space each term dominates.
There are four terms in the partition function, corresponding to clusters that are regular ($\Delta B = 0$, $d_X=0$), singular isostatic ($\Delta B = 0$, $d_X=1$), hyperstatic ($\Delta B = 1$, $d_X=0$), and hypostatic ($\Delta B = -1$, $d_X=2$). When $\gamma$ is large enough and $\alpha$ is fixed or grows more slowly than $\gamma$,
the hyperstatic cluster dominates. We also see that for fixed $\gamma$ and large enough $\alpha$, the hypostatic cluster appears to dominate. However, in that
region of parameter space, the sticky parameter $\kappa$ tends to zero and we do not expect the spheres to leave the fluid phase and stick to each other.
 Figure \ref{fig:region1} shows lines of constant $\kappa$ for $\kappa = 10$ and $1000$; these are values between which we could reasonably expect to see clusters that equilibrate over experimental timescales, rather than a gas or a glass.
 Figure \ref{fig:region1} also shows the 
relative frequencies as a function of $\alpha$ for $\kappa=40$, a reasonable value to observe in experiments \cite{Perry:2015ku}, where we see the transition from the region
dominated by regular clusters, favored by the higher number of distinct clusters, to that dominated by the hyperstatic cluster, favored by lower energy.
The singular clusters peak in frequency near this transition, but never rise above 3.6\%. The hypostatic cluster is extremely rare and never reaches
above 0.03\% despite its high vibrational entropy.

For $N=11$ to $14$, we use the enumeration of rigid clusters to construct similar phase diagrams, shown in  Figure \ref{fig:region2}. Again, in the region of $\kappa$ where
we can expect to observe the formation of equilibrated clusters, we primarily see a transition between regular clusters and the most hyperstatic ones.
The exception is $N=13$, for which two of the eight most hyperstatic clusters are singular and dominate in the sticky limit.

For $N=15$ to $19$, the maximum value of $\Delta B$ is $5,6,7,8,$ and $9$ respectively, and only clusters with at least 4 bonds less than the maximum were enumerated
\cite{HolmesCerfon:2016wa}. The phase diagrams computed are therefore not a complete description of rigid clusters, but their common appearance and patterns
is a telling signal of emergence. In particular, as in most previous cases, at fixed $\kappa\gtrsim 1$ and increasing $\alpha$, we observe a transition from
dominance of the least hyperstatic nonsingular clusters among those enumerated to dominance of the most hyperstatic. For $N=11$ to $14$, the former are regular clusters,
but regular clusters are too numerous to have been enumerated for $N\ge15$. This transition occurs in a narrow range of $\gamma$, with possibly a few intermediate
regions.

\section{Discussion}

\subsection{Toward larger $N$}
Our observations about the free-energy landscape of clusters of up to $19$ sticky spheres suggest two conjectures about
the landscape as $N$ increases: 
\begin{enumerate}
\item For most values of $N$, singular clusters are rare at experimentally relevant values of $\kappa$;
\item The phase diagrams of rigid clusters have a universal shape, which is approached even at small $N$. 
\end{enumerate}

The first conjecture, if true, goes much of the way toward explaining the phenomenon of emergence.
It implies that despite the apparent competition between the higher vibrational entropy of singular clusters
and the lower energy of hyperstatic clusters, the latter always wins. 
Since the most hyperstatic clusters are close to fragments of a close-packed lattice with defects only on the surface, close-packing order
should arise for sticky spheres even for small clusters. 

Based on \eqref{eq:Zfinal}, we have already noted that each extra bond contributes
to the stability of a rigid cluster by a factor $\gamma$, and each singular mode contributes by a factor $\alpha$.
This surmise neglects the geometric factor, which might be correlated with the number of bonds or singular modes,
and we return to such correlations shortly. Since
$\gamma\propto\kappa\alpha^2$, and assuming that $\kappa\gtrsim1$, it follows that as $\alpha\to\infty$, each extra
bond contributes  \emph{at least} twice as much as each singular mode. 
In this sense, the observation of Meng et al.\ that a zero mode contributes to stability about half as much as an extra bond turns out to be rather prescient,
even as it was surmised from experiments far from the Baxter limit \cite{Meng2010}.

Therefore, the only way
a rigid cluster with fewer than the maximum number of bonds can become dominant at large $\alpha$ is if it had two
singular modes for each missing bond. This scenario does not seem to occur for any value of $N$.
In fact, the only exceptions to the rarity of singular clusters are when singular clusters exist among
the most hyperstatic ones, as they do for $N=9$ and $13$. We do not know if such clusters might exist for larger values of $N$.
Such ``magic numbers,'' associated with special clusters and deviating from the trend toward bulk behavior, are familiar in other systems, 
such as atomic clusters \cite{Echt:1981bw,Raoult:1989dl,Wales:1997jj,Wales:2012dd}.

\begin{figure*}
\includegraphics[trim={1cm 0 4cm 0},clip,width=0.9\linewidth]{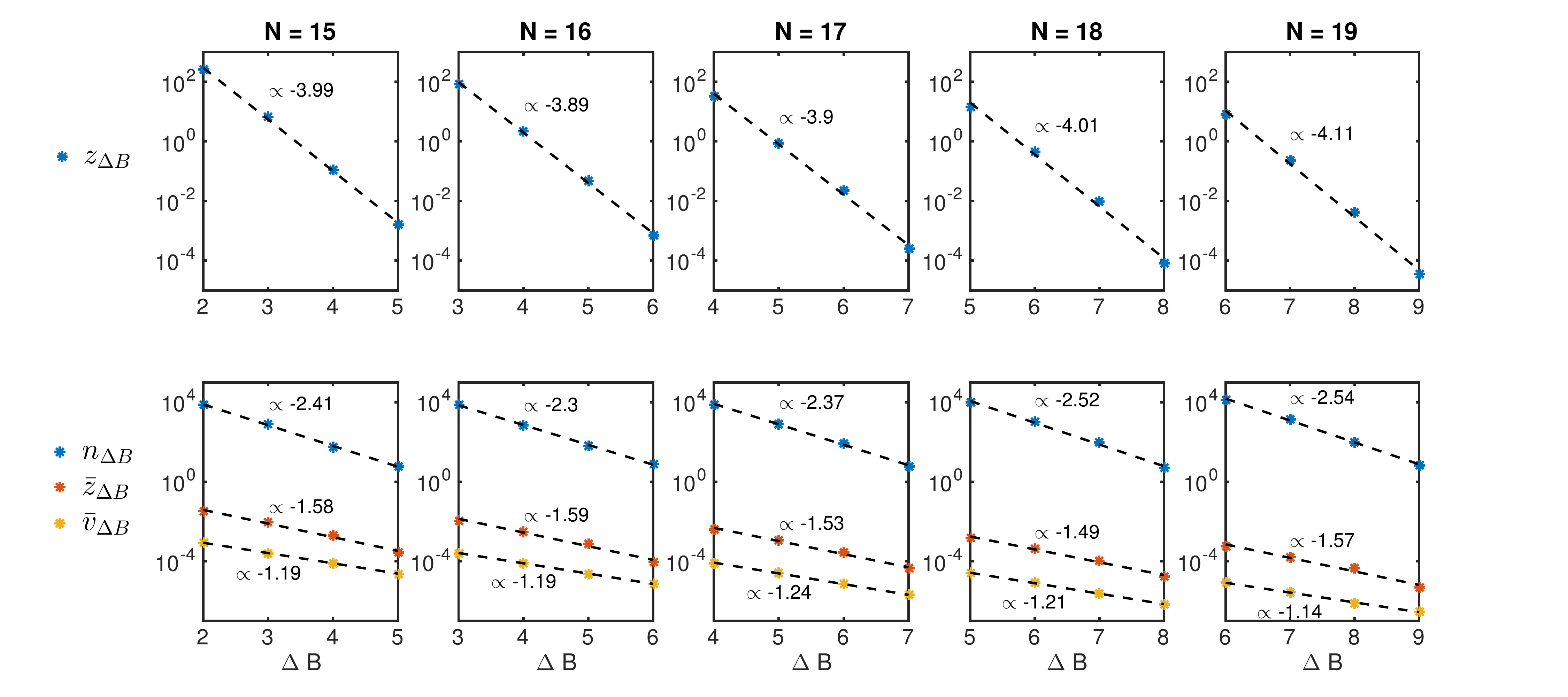}
\caption{Top row: the total geometric partition function of rigid clusters with $\Delta B$ bonds above the isostatic number of bonds as a function of $\Delta B$. The dashed
    line is an exponential fit, $\log z_{\Delta B} = c_1 + c_2 \Delta B$, with the best fit value of $c_2$ indicated. The bottom row shows different contributions to
    the value of $z_{\Delta B}$: $n_{\Delta B}$ is the number of distinct rigid clusters, $\bar z_{\Delta B}$ is the mean geometric partition function of such a cluster,
    and $\bar v_{\Delta B}$ is the mean vibrational contribution to the latter. 
    \label{fig:scalings}}
\end{figure*}

Motivated by the first conjecture, we limit our attention to nonsingular clusters and consider the second conjecture. The transition from a region dominated by
regular clusters to one dominated by hyperstatic ones occurs not perfectly abruptly but nevertheless over a small range of $\gamma$. Such behavior
implies that the total geometric partition function $z_{\Delta B}$ (we drop the second subscript for the remainder of the section,
implicitly taking $d_X=0$) decreases exponentially as a function of $\Delta B$ with a roughly consistent rate
$z_{\Delta B}\sim \exp(-\lambda \Delta B)$, as can in fact be seen in  Figure \ref{fig:scalings}. Moreover, the rate is consistent also across $N$, lying between $\lambda=3.9$
and $\lambda =4.1$ for $15\le N\le19$ with no pronounced trend. Therefore, the center of the transition, at $\gamma=\exp(\lambda)$, is also consistent across $N$.

Which factor in the partition function contributes most strongly to this exponential scaling?
Part of it is due to the number of distinct clusters with a given number of bonds, $n_{\Delta B}$, whose logarithm can be thought of as a
configurational entropy.  Figure \ref{fig:scalings} shows it decreases exponentially with $\Delta B$. 
The average geometric partition function 
of a typical single cluster with a certain number of bonds 
also appears to decrease exponentially as a function of $\Delta B$, with most
of this decreases accounted for by the decrease in the average vibrational factor $(\det \tilde{M})^{-1/2}$ per cluster. 
The deviations from the exponential fits for these individual components are more pronounced than in the case of the total geometric partition function, but the consistent exponential scaling is notable. 
The rotational contributions, from the moment of inertia and symmetry factors, do not depend strongly on the number of bonds (SI Section \ref{sec:scalings}.)

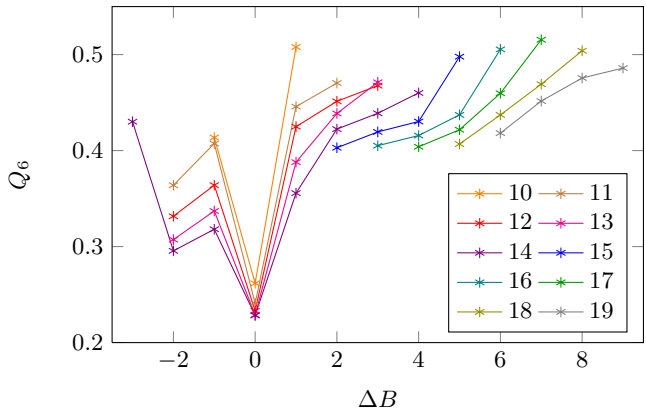
\begin{figure}
    \begin{center}
   	\begin{tikzpicture}
	    \begin{axis}[
		    width=\linewidth,height=0.7\linewidth,
		    xlabel={$\Delta B$}, ylabel={$Q_6$},
		    xmin=-3.5, xmax=9.5,
		    ymin=0.2, ymax=0.55,
		    domain=-3.5:9.5,
		    legend style={at={(0.97,0.03)},anchor=south east},
		    legend entries={10,11,12,13,14,15,16,17,18,19},
		    legend columns=2,
		    yticklabel style = { /pgf/number format/fixed },
		]
	        \addplot[mark=asterisk,color=orange] table[x index=0, y index=1] {q6-n10.dat};
	        \addplot[mark=asterisk,color=brown] table[x index=0, y index=1] {q6-n11.dat};
	        \addplot[mark=asterisk,color=red] table[x index=0, y index=1] {q6-n12.dat};
	        \addplot[mark=asterisk,color=magenta] table[x index=0, y index=1] {q6-n13.dat};
	        \addplot[mark=asterisk,color=violet] table[x index=0, y index=1] {q6-n14.dat};
	        \addplot[mark=asterisk,color=blue] table[x index=0, y index=1] {q6-n15.dat};
	        \addplot[mark=asterisk,color=teal] table[x index=0, y index=1] {q6-n16.dat};
	        \addplot[mark=asterisk,color=green!60!black] table[x index=0, y index=1] {q6-n17.dat};
	        \addplot[mark=asterisk,color=olive] table[x index=0, y index=1] {q6-n18.dat};
	        \addplot[mark=asterisk,color=gray] table[x index=0, y index=1] {q6-n19.dat};
	    \end{axis}
	\end{tikzpicture}
    \end{center}
\caption{Mean value of the bond-orientational order parameter $Q_6$ as a function
    of the number of extra bonds $\Delta B$. The mean includes all clusters, regular and singular, weighted uniformly. 
    \label{fig:q6}}
\end{figure}

To further investigate how these observations bear on the question of emergence of long range order, we consider a metric for the bond-orientational order. We use the order metric $Q_6$, which is widely used in the study of the structure of liquids (see SI Section \ref{sec:q6} for a definition) \cite{Steinhardt1981}.
 Figure \ref{fig:q6} shows that bond-orientational order is strongly correlated with the existence and number of extra bonds. Isostatic clusters have
particularly low bond-orientation order. Deviation from isostaticity in either direction seems to require the introduction of partial order, since only when
going away from the generic situation can the algebraic singularities associated with hypostaticity and hyperstaticity arise.
The presence of even a single extra bond seems to require the structure to contain two octahedra sharing an edge, and as more
bonds are added, this substructure can grow to incorporate larger and larger close-packing fragments.
The most hyperstatic clusters observed for each $N$ are fragments of close-packing
lattices with defects only on the surface.

\begin{figure*}
\begin{center}
\includegraphics[trim={0cm 0 0cm 0},clip,width=\linewidth]{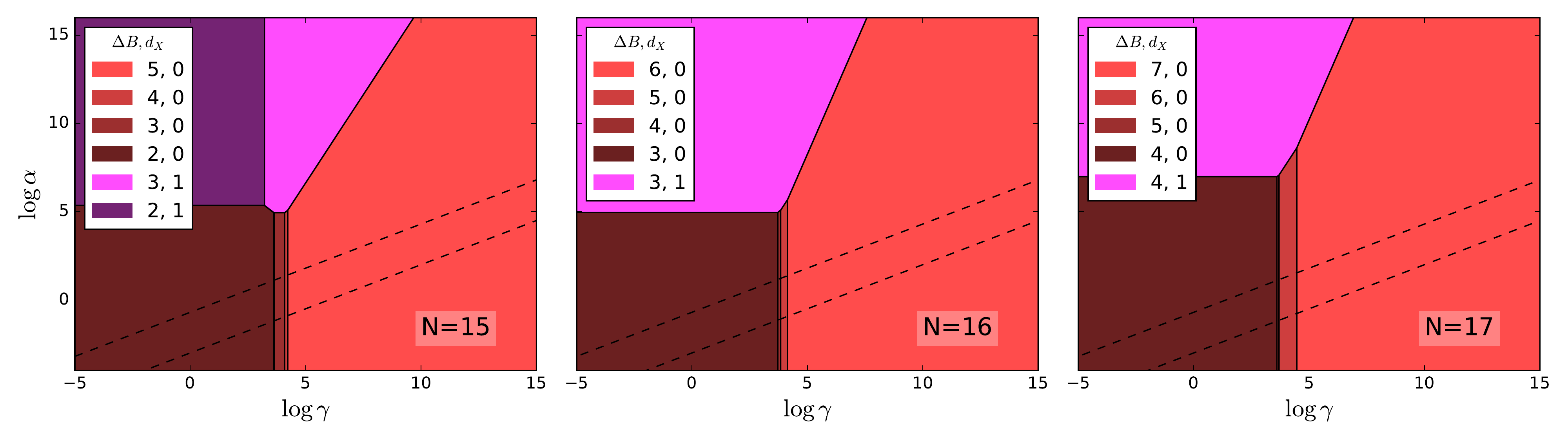}\\
\includegraphics[trim={0cm 0 0cm 0},clip,width=0.67\linewidth]{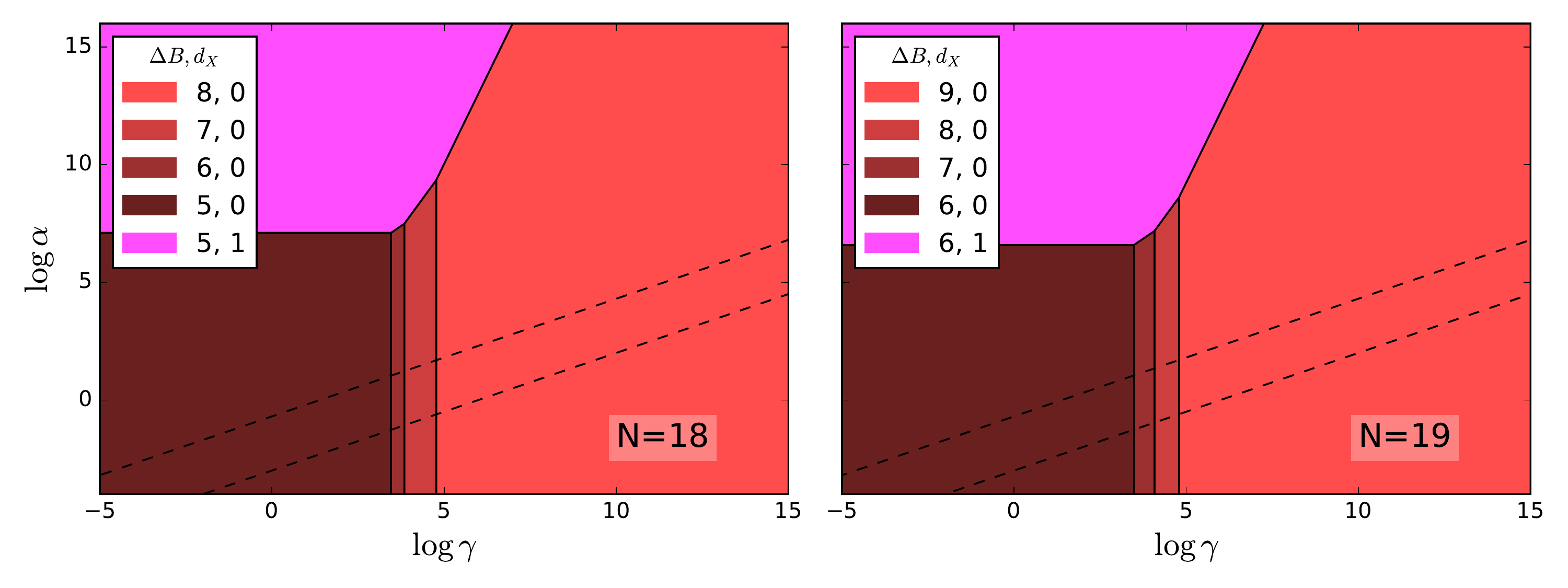}
\end{center}
\caption{Phase diagrams for rigid clusters of $15$--$19$ spheres.
Labeling is the same as in  Figure \ref{fig:region1}. 
\label{fig:region3}}
\end{figure*}

\subsection{Relation to bulk behavior}

In addition to being physically important for their own sake, finite clusters also carry importance
as a formal tool in low-density expansions of the fluid behavior of bulk systems. Baxter used the expansion
up to the second term to predict a gas-liquid transition in a fluid of sticky spheres \cite{Baxter1968}.
It has been previously observed that Baxter's calculation is theoretically problematic because later
terms in the expansion diverge in Baxter's limit \cite{Stell1991}.
Specifically, Stell and Williams showed that for $N=12$ spheres the partition function
diverges, due to the existence of rigid clusters which remain rigid even when
turning off the interaction between some bonded pair \cite{Stell1991}. This divergence in fact occurs for
any hyperstatic cluster, leading to divergence even for $N=10$. Here, we have shown
that singular clusters also lead to a divergence in Baxter's limit and so the theoretical problems for the
expansion begin even earlier, at $N=9$.

Our numerical results show a sharp transition to the prevalence of isostatic structures
over highly hyperstatic, close-packed ones when the depth of the potential well is less
than about $4k_B T$, and we expect it to extend into the bulk regime.

\section{Conclusions and Future Work}

We have shown that the free energies of systems of spheres interacting with a very short-ranged pair potential
can be described by only two parameters, characterizing the depth and width of the potential. This is true even
when the system is in a singular energy minimum, where the harmonic
approximation diverges, provided the associated framework is second-order rigid.
We used our results to study the free energy landscapes of clusters of a finite number $N\leq 19$ of spheres. Previous work has shown that for $N\leq 8$,
the most likely clusters to be observed in equilibrium are the least symmetric ones.  We showed that for larger $N$,
the free energy landscape is dominated by the most hyperstatic clusters. These clusters are usually close to fragments of a close-packed lattice with defects only on the surface, so we see the emergence of crystalline order even for $N$ as small as 10.
Only for the magic numbers, $N=9$ and $N=13$, is the maximum number of contacts achieved by both nonsingular and singular clusters, and the latter
would form with near unit probability in the limit of zero range interactions.

Our calculation is different from simply including all anharmonic corrections to the free energy, as has been done numerically
in molecular dynamics studies of Lennard-Jones clusters \cite{Doye1995Anharmonic}.
Calculating these is usually a Sisyphean task, requiring heavy lifting and providing only minor corrections.
However, when the vibrational spectrum contains zero modes, these corrections become crucial. Thankfully, as we have shown, the anharmonic corrections need
only be calculated for the zero modes, though care must be taken to account for the coupling between the zero and nonzero modes, which has
previously been neglected \cite{Meng2010}. 

One possible issue with using our result is the evaluation of the integral $\int_{\mathbb{R}^n} \exp[-Q(\bb{x})]d\bb{x}$,
where $Q$ is a positive quartic form. Unlike its Gaussian relative, this integral has no general analytic solution
for even moderate values of $n$ \cite{Morozov2009}. A future avenue of research is to simplify the
integral under stronger rigidity assumptions, such as prestress stability.

We found empirically that the partition function of rigid clusters of size $N$ with $B$ contacts scales exponentially with $B$. If this scaling continues, it leads to a sharp transition between a region of parameter space where clusters with the largest possible number of contacts dominate and a region
where the expected number of contacts is extensively less than the maximum.
This sharp transition may be a finite-size incarnation of the thermodynamic observation that particles with a short-range interaction have little or no liquid phase, a fact that has made it hard to create liquids of biomolecules or other nanoscale objects \cite{Perriman:2011bq}.
We analyzed the factors contributing to this scaling, but do not have a explanation from first principles that would
suggest the scaling continues for larger sizes. We leave this as an open problem that could be addressed using ideas from geometry and statistical mechanics. 

We have focused our calculations on rigid clusters, but one could use \eqref{eq:floppy} to compute the partition functions of floppy clusters or other networks with floppy modes of deformation \cite[e.g.][]{Chen:2014ec}. 
This is more challenging because it involves computing integrals over manifolds, but is still tractable using specialized parameterizations as in \cite{Ozkan:2011vy,Holmes2013} or thermodynamic integration techniques \cite{Frenkel:2001uy}. Moreover, current numerical and analytic methods for classifying the rigidity of bar frameworks focus on testing
rigidity, not floppiness, and new methods are required for determining what subspace of the space of infinitesimal flexes (zero modes) extends to finite degrees of freedom in a floppy cluster.

We showed via simulations that the sticky limit describes real, finite-range potentials, provided the range of the potential is much shorter than minimum gap between non-contacting spheres. However, this minimum gap appears to become arbitrarily small as $N$ increases \cite{HolmesCerfon:2016wa}. Nevertheless, we  expect our calculations to give qualitative insights into real systems and it would be interesting to test them with experiments on clusters larger than $N=9$. For example it could still be the case that the
type of clusters that dominate in equilibrium depends on only a small number of parameters as the interaction potential is varied, or that the ones that are close to singular become more probable as the range of the potential is decreased. 

Although the gap size poses a problem for quantitative agreements in real systems, it may be possible to build on the present results by including perturbations that account for non-nearest neighbor interactions.
As non-nearest neighbor interactions increase, the energy minima coming from distinct rigid clusters are expected to merge, leading to
critical points that can be analyzed using  catastrophe theory \cite{Wales2001} or related ideas. 
Singular clusters themselves can be
thought of as multiple, colocated minima that
might separate as they are perturbed by non-nearest neighbor interactions. 
By starting with the sticky limit and slowly increasing the range of the interaction potential, one might create a bifurcation diagram of energy landscapes which shows a finite or low-dimensional number of possible landscapes; indeed in related systems it has been shown that the space of energy-minimizing configurations is sometimes much lower-dimensional than the space of interaction potentials \cite{Ballinger:2009jj}. 

Many systems beyond clusters, particularly metamaterials and systems near jamming or close packing, are often modeled as subject to soft constraints
in the limit that the stiffness of the constraint is taken to infinity. As we have shown here, this limit is qualitatively different
than modeling the constraints as hard constraints, particularly when the excitation spectrum includes zero modes. In this paper, we have
focused on the thermodynamic effect of zero modes, calculating their contribution to the free energy. Such thermal effects are already
observed in many metamaterials and self-assembly systems, and are likely to become important for more of them as they are fabricated
on smaller scales. The zero-temperature effects of zero modes are also potentially important in metamaterials and athermal fluids and
more work will need to be done to determine their effect on mechanical stability and yielding behavior.
We hope that the present work will become the first step in a broader program to provide analytic tools for going beyond the harmonic
approximation and calculating the thermodynamic and mechanical effects of zero-frequency vibration modes.





\bigskip

\begin{acknowledgments}
Thanks to Michael O'Neil for helpful discussions. 
Y.\ K.\ was supported by an Omidyar Fellowship at the Santa Fe Institute.
M.\ H.-C.\ was supported by the U.S. Department of Energy, Office of Science, Office of Advanced Scientific Computing Research under award DE-SC0012296.
\end{acknowledgments}

\bibliographystyle{apsrev4-1}
\bibliography{landscape.bib,RefsClusters.bib}



\appendix

\section{Details on the partition function calculations}\label{sec:calcs}

\subsection{Sticky limit}

Starting from an arbitrary potential $V_0(r)$ we can define a one-parameter family of rescaled potentials,
parameterized by the variable $\epsilon$, representing the interaction range.
Let $V(r) = C(\eps) V_0( (r-d)d/\eps + d)$. We need to extend $V_0$ for negative arguments
in some way such that $V_0(r) \to \infty$ as $r\to-\infty$. Assuming such scaling is used to approach
the sticky limit, then all derivatives of $V(r)$ at $r=d$ satisfy the scaling $d^nV(r)/dr^n|_{r=d} = O(\mathcal E/\eps^{n})$,
where $\mathcal E = -V(d)$. We assume this scaling of the derivatives for deriving all the big-O error bounds in
our calculation.

The traditional sticky sphere limit of Baxter \cite{Baxter1968} is defined so that the difference between the partition function for two spheres interacting
via $V(r)$ and the partition function for two hard spheres approaches a constant. Namely,
\begin{equation}
    0 < \lim_{\eps\to0} \frac{1}{d^3}\int_{\mathbf{r}\in\mathbb{R}^3}(e^{-\beta V(|\mathbf{r}|)}-1)d^3\bb{r}+\frac{4\pi}3 < \infty\text,
\end{equation}
where $\beta=(k_BT)^{-1}$ is the inverse of temperature $T$ times the Boltzmann constant.
The limit can be evaluated using Laplace's method as
\begin{equation}
    2^{5/2}\pi^{3/2}\lim_{\eps\to 0} (a \beta d^2)^{-1/2}\exp(\beta \mathcal E)\text,
\end{equation}
where $\mathcal E = -V(d) = -C(\eps)V_0(d)$ and
$a=V''(d) = C(\eps)V_0''(d)d^2/\eps^2$.

In our calculation we do not assume this limit, letting $\eps$ and $C$ approach $0$ and $\infty$
simultaneously in any way desirable. We define the sticky parameter to be 
$\kappa = (2\pi)^{1/2} (a \beta d^2)^{-1/2}\exp(\beta \mathcal E) =  (2\pi)^{1/2} \alpha^{-2} \gamma$ even
when the limit is not taken.

\begin{widetext}

\subsection{Expansion of the pairwise potential}

We consider the Taylor expansion of $V(|\bb{r}_{ij} + \delta \bb{r}_{ij}|)$ in powers of $\delta \bb{r}_{ij}$,
where $|\mathbf{r}_{ij}|=d$. Applying the chain rule and using the Taylor expansion of the square root,
we get
\begin{equation}
    \begin{aligned}
	V(|\bb{r}_{ij} + \delta \bb{r}_{ij}|) &= V(\sqrt{ d^2 + |\delta\bb{r}_{ij}|^2 + 2 \mathbf{r}_{ij}\cdot\delta\mathbf{r}_{ij}})\\
	&= \sum_{l=0}^\infty \frac{1}{l!}V^{(l)}(d)
	\left[\sum_{k=0}^\infty {1/2 \choose k} d^{1-2k} (|\delta\bb{r}_{ij}|^2 + 2 \mathbf{r}_{ij}\cdot\delta\mathbf{r}_{ij})^k\right]^l\text,
    \end{aligned}
\end{equation}
where ${1/2 \choose k} = \tfrac1{k!}\prod_{n=0}^{k-1} (\tfrac12 - n)$ is the generalized binomial coefficient.

When we expand the sum and keep terms up to fourth degree in $\delta \bb{r}_{ij}$, recalling that $V'(d)=0$, we get
\begin{equation}\label{eq:SIexpand}
    \begin{aligned}
	V(\|\mathbf{r}_{ij}+\delta\mathbf{r}_{ij}\|) =& -\mathcal{E} +
	\frac{a}{2d^2} (\mathbf{r}_{ij}\cdot\delta\mathbf{r}_{ij})^2 +
	\frac{b}{6d^3}(\mathbf{r}_{ij}\cdot\delta\mathbf{r}_{ij})^3 +\\
	&\frac{a}{2d^2}(\mathbf{r}_{ij}\cdot\delta\mathbf{r}_{ij})(\delta\mathbf{r}_{ij}\cdot\delta\mathbf{r}_{ij}) + 
	\frac{c}{24d^4}(\mathbf{r}_{ij}\cdot\delta\mathbf{r}_{ij})^4 +\\
	&\frac{b}{4d^3}(\mathbf{r}_{ij}\cdot\delta\mathbf{r}_{ij})^2(\delta\mathbf{r}_{ij}\cdot\delta\mathbf{r}_{ij}) +
	\frac{a}{8d^2}(\delta\mathbf{r}_{ij}\cdot\delta\mathbf{r}_{ij})^2 + \\ & O(\mathcal E \|\delta \mathbf{r}_{ij}\|^5/\eps^5)\text,
    \end{aligned}
\end{equation}
where $\mathcal E = -V(d)$, $a=V''(d)$, $b=V'''(d)-3V''(d)/d$, $c=V^{(4)}(d)-6V'''(d)/d+15V''(d)/d^2$.

Later on in the calculation, we decompose the displacement into orthogonal components $\delta\bb{r} = \bb{x}+\bb{y}$ with the property
that $\mathbf{r}_{ij}\cdot\mathbf{x}_{ij} = 0$ for all $(i,j)\in E$. When we plug this decomposition into Equation \ref{eq:SIexpand} and
keep terms below $O(\tfrac{\mathcal E}{\eps^3 d^3}(\|\mathbf{x}\|^2+d\|\mathbf{y}\|)^3)$, we get
\begin{equation}\label{eq:vij-xy}
    \begin{aligned}
	V(|\mathbf{r}_{ij}+\mathbf{x}_{ij}+\mathbf{y}_{ij}|) =&  -\mathcal{E} +
	\frac{a}{8d^2}(\mathbf{x}_{ij}\cdot\mathbf{x}_{ij})^2 +\\
	&\frac{a}{2d^2}(\mathbf{x}_{ij}\cdot\mathbf{x}_{ij})(\mathbf{r}_{ij}\cdot\mathbf{y}_{ij}) +
	\frac{a}{2d^2}(\mathbf{x}_{ij}\cdot\mathbf{x}_{ij})(\mathbf{x}_{ij}\cdot\mathbf{y}_{ij})+\\
	&\frac{a}{2d^2}(\mathbf{r}_{ij}\cdot\mathbf{y}_{ij})^2 +
	\frac{a}{d^2}(\mathbf{x}_{ij}\cdot\mathbf{y}_{ij})(\mathbf{r}_{ij}\cdot\mathbf{y}_{ij})+\\
	&\frac{a}{4d^2}(\mathbf{x}_{ij}\cdot\mathbf{x}_{ij})(\mathbf{y}_{ij}\cdot\mathbf{y}_{ij})+
	\frac{a}{2d^2}(\mathbf{x}_{ij}\cdot\mathbf{y}_{ij})^2+\\
	&
	O(\tfrac{\mathcal E}{\eps^3 d^3}(\|\mathbf{x}\|^2+d\|\mathbf{y}\|)^3)\text.
    \end{aligned}
\end{equation}

\subsection{Free energy of a rigid cluster}

Consider the coordinates of the centers of $N$ spheres, $\mathbf{r}_i\in\R^3$, $i=1,\ldots,N$, which we consider as a single
point $\mathbf{r}=\mathbf{r}_1\oplus\ldots\oplus\mathbf{r}_N\in\R^{3N}$ of the configuration space. Let
$E = \{ (i_1,j_1),\ldots,(i_B,j_B)\}$ be the complete list of pairs such that
\begin{equation}\label{eq:againbonds}
    |\bb{r}_{i} - \bb{r}_{j}|^2 = d^2, \qquad (i,j)\in E\text.
\end{equation}
We say that the point $\mathbf{r}$ describes a rigid cluster if any other point in the connected component
of the solution space to \ref{eq:againbonds} that contains $\mathbf{r}$ is related to $\mathbf{r}$ by translation and rotation.

\paragraph{Partition function}

We ignore momenta in our phase space because they contribute a constant term to the partition function. Therefore,
the partition function is defined as 
\begin{equation}
    Z = \frac{1}{N!}\int_{\mathbf{r}'\in \Omega^n} \exp [ -\beta U(\mathbf{r}') ]d\mathbf{r}'\text,
\end{equation}
where $\Omega$ is the region of space to which the particles are confined, and has volume $V$.

\paragraph{Domain of integration}

The free energy of a particular rigid cluster $\mathbf{r}$ is defined to be $-\beta \log Z_\mathbf{r}$,
where $Z_\mathbf{r}$ is the contribution to the total partition function due to the configurations that
are associated with the rigid cluster $\mathbf{r}$. The space of such configurations, denoted $\mathcal N_\mathbf{r}$ is
defined to be the union over all permutations $\rho\in S_N$ of domains of the form
\begin{equation}
    \{\bb{r}'\in \Omega^{N}: \mathbf{r}'_i = U(\bb{r}_{\rho(i)}) + \mathbf{t} + \delta \mathbf{r}_i \text{ for some } U\in O(3), \mathbf{t}\in\R^3,
    \mathbf{\delta r}\in \R^{3N} \text{ such that }|\delta\mathbf{r}_i|\le l \}\text,
\end{equation}
where $l$ is chosen such that $\eps\ll l \ll d$.

\paragraph{Change of variables}

We can remove the rigid-body degrees of freedom and the permutations from the calculation
by performing a change of variables.
Fix $\mathbf{r}$ to be some configuration corresponding to the cluster of interest.
Let $W=\{\delta\mathbf{r}\in\mathbb{R}^{3N} : \delta\mathbf{r}_i = \mathbf{t} + \mathbf{s}\times\mathbf{r}_i \text{ for some } \mathbf{t}, \mathbf{s}\}\subseteq \R^{3N}$,
and let $W^\perp$ be its orthogonal complement as a linear subspace of $\mathbb{R}^{3N}$. The transformation 
$\phi\colon W^\perp\times\mathbf{R}^3\times SO(3) \to \mathbf{R}^{3N}$ that maps
$\phi\colon(\delta\mathbf{r}_i,\mathbf{t},U)\mapsto U(\mathbf{r}_i +\delta \mathbf{r}_i) + \mathbf{t} - \mathbf{r}_i$
is nonsingular and its Jacobian determinant is
\begin{equation}
    J_\phi(\delta\mathbf{r}) = I + O(d^2\|\delta\mathbf{r}\|)\text,
\end{equation}
where $I$ 
is the square root of the determinant of the moment of inertia tensor about the center of mass.
Summing over permutations contributes a factor of $N!/\sigma$, where $\sigma$ is the order of
the symmetry group of the cluster, that is, the number of Euclidean motions (combinations of reflections, rotations and translations)
that map $\mathbf{r}$ to a permutation of itself. In the new variables, the integrand is independent of $U$ and $\mathbf{t}$, and we can
integrate those variables. The only complication is that the range over which $\mathbf{t}$ varies, the free volume available to
the cluster, can be slightly different for different clusters due to boundary effects. However, these differences are of order $V^{2/3}d$,
compared to the total volume $V$, so they are negligible if $V\gg d^3$.

Following the change of variables and the integration over the rigid body degrees of freedom, $Z_\mathbf{r}$ reduces to 
\begin{equation}\label{eq:SIZe2p}
    Z_\mathbf{r}=\frac{V}{\sigma}\int_{\delta\mathbf{r}\in W^\perp,|\delta\mathbf{r}_i|\le l} \exp\left[-\beta U(\mathbf{r}+\delta\mathbf{r})\right] J_\phi(\delta\mathbf{r})d(\delta\mathbf{r})\text,
\end{equation}
where we have removed a factor of $N!$ and the dimensionless factor obtained from integrating over $SO(3)$.
The inclusion of $N!$ depends on whether we wish to treat the spheres as distinguishable or not, but in either case
it will not change the results as it is common to all clusters. The free energy is given by $F_\mathbf{r} = -\beta^{-1}\log Z_\mathbf{r}$. 
Finally, we can replace $U(\mathbf{r}+\delta\mathbf{r})$ by $U_E(\mathbf{r}+\delta\mathbf{r})=\sum_{(i,j)\in E} V_{ij}(\mathbf{r}_{ij}+\delta\mathbf{r}_{ij})$
and extend the domain of integration to the entire linear space $W^\perp$:
\begin{equation}\label{eq:SIZe2}
    Z_\mathbf{r}=\frac{V}{\sigma}\int_{\delta\mathbf{r}\in W^\perp} \exp\left[-\beta U_E(\mathbf{r}+\delta\mathbf{r})\right] J_\phi(\delta\mathbf{r})d(\delta\mathbf{r})\text.
\end{equation}

\paragraph{Harmonic approximation}
To evaluate the integral in \eqref{eq:SIZe2}, we use the expansion \ref{eq:SIexpand}.
Keeping terms up to second order in the displacements yields
\begin{equation}\label{eq:SIU2}
    U_E(\mathbf{r}+\delta\mathbf{r}) = -B\mathcal{E} + \tfrac12\langle\delta\mathbf{r},M \delta\mathbf{r}\rangle+O(\mathcal E\|\delta\mathbf{r}\|^3/\epsilon^3)\text,
\end{equation}
where $M$ is a symmetric linear map, whose matrix representation is usually referred to as the \emph{dynamical matrix}.
When \eqref{eq:SIU2} is plugged into the integral in \eqref{eq:SIZe2} without the error term, the integral converges if and only if $M$,
considered as a map $W^\perp\to W^\perp$ is positive definite, that is, all its eigenvalues are positive. The result
gives the harmonic approximation for the vibrational partition function and the error term contributes a multiplicative
correction of the form $1+O(\beta^{-1/2} \mathcal E^{-1/2})$.

\paragraph{Fourth-order approximation}
The harmonic approximation fails when the null space of the dynamical matrix extends beyond $W$,
or equivalently when nontrivial flexes of the rigidity matrix exist. These are infinitesimal degrees of freedom that,
if the cluster is rigid, are not
extendable to finite degrees of freedom. We call them \emph{singular} directions. 
Let $X=\mathrm{ns} (M)\cap W^\perp$ and $Y=(\mathrm{ns} (M))^\perp$
so that $W^\perp = X\oplus Y$ is an orthogonal decomposition of $W^\perp$,
and $X$ is the null space of $M$ restricted to $W^\perp$.
For every $\delta \bb{r} \in W^\perp$ we may write $\delta \bb{r} = \bb{x} + \bb{y}$ where $\bb{x} \in X$, $\bb{y}\in Y$.
Using the expansion \ref{eq:vij-xy}, we get
\begin{equation}\label{U4}
    U(\mathbf{r}+\mathbf{x}+\mathbf{y}) = -B\mathcal{E}+U_0(\mathbf{x}) +
                                          \langle\mathbf{u}_1(\mathbf{x}),\mathbf{y}\rangle + 
					  \tfrac12\langle\mathbf{y} , U_2(\mathbf{x})\mathbf{y}\rangle +
					  O(\mathcal E\|\mathbf{y}\|^3/\epsilon^3)\text,
\end{equation}
Here $U_0(\bb{x})$ is a real scalar, $\bb{u}_1(\bb{x})\in Y$ is a $3N$-dimensional vector, and $U_2(\bb{x}):Y\to Y$ is a real symmetric map. These have the explicit forms
\begin{align}
    U_0(\bb{x}) &= \sum_{(ij)\in E} \frac{a}{8d^2} (\mathbf{x}_{ij}\cdot\mathbf{x}_{ij})^2 + O(\mathcal E \|x\|^6/d^3\eps^3)\\
    [\mathbf{u}_1(\bb{x})]_i &= \sum_{j\text{ s.t. }(i,j)\in E} \frac{a}{2d^2} (\mathbf{x}_{ij}\cdot\mathbf{x}_{ij}) \mathbf{r}_{ij} + O(\mathcal{E} \|x\|^3/d^2\eps^2 + \mathcal{E} \|x\|^4/d^2\eps^3)\\
U_2(\bb{x}) &= M + O(\mathcal E \|\bb{x}\|^2/d\epsilon^3)
\end{align}
We write $U_0(\mathbf{x}) = a d^2\tilde{U}_0(\mathbf{x}) + O(\mathcal E \|x\|^6/d^3\eps^3)$,
$\mathbf{u}_1(\mathbf{x}) = ad\tilde{\mathbf{u}}_1(\mathbf{x})+O(\mathcal{E} \|x\|^3/d^2\eps^2 + \mathcal{E} \|x\|^4/d^2\eps^3)$,
and $U_2(\mathbf{x}) = a \tilde M + O(\mathcal E \|\bb{x}\|^2/d\epsilon^3)$, where $\tilde{U}_0(\mathbf{x})$ is a homogeneous
quartic function of $\mathbf{x}$, $\tilde{\mathbf{u}}_1(\mathbf{x})$ is a homogeneous quadratic function of $\mathbf{x}$, and
$\tilde M = M/a$ is the geometric part of the dynamical matrix. Let
\begin{equation}
    \tilde U(\mathbf{x},\mathbf{y}) = \tilde{U}_0(\mathbf{x}) + 
                                      \frac{1}{d}\langle\tilde{\mathbf{u}}_1(\mathbf{x}),\mathbf{y}\rangle + 
				      \frac{1}{2d^2}\langle\mathbf{y},\tilde{M} \mathbf{y})\text.
\end{equation}
As discussed in the text, $\tilde{U}(\bb{x},\bb{y})$ is positive for all nonzero $(\bb{x},\bb{y})$ if and only if the cluster \emph{second-order rigid}.

In this case, we can calculate
the leading-order term of the partition function. We first integrate over $Y$ by completing the square
and computing the simple Gaussian integral. The remaining integral over $X$ is nontrivial \cite{Morozov2009},
and we leave it unevaluated, but dimensionally reduce it:
\begin{equation}\label{eq:SIintx}
    \begin{aligned}
	Z&=\frac{ I V e^{\beta B\mathcal{E}}}{\sigma}\int_{X}d\mathbf{x}\int_{Y}d\mathbf{y} \exp\left[
	    -a\beta d^2 \tilde U(\mathbf{x},\mathbf{y}) + O(\epsilon_1)
        \right]\\
	&=\frac{ I V e^{\beta B\mathcal{E}}}{(\det \tilde{M}|_Y)^{1/2}\sigma} \left(\frac{2\pi}{a\beta}\right)^{\frac{d_Y}{2}}
	\int_{X}d\mathbf{x} \exp\left[-a \beta d^2 \tilde{U}_\text{min}(\mathbf{x})+O(\epsilon_2)\right]\\
	&=\frac{ I V e^{\beta B\mathcal{E}}}{(\det \tilde{M}|_Y)^{1/2}\sigma} \left(\frac{2\pi}{a\beta}\right)^{\frac{d_Y}{2}}
	\left(\frac{d^2}{a\beta}\right)^{\frac{d_X}{4}} \left(\int_{X}e^{-Q(\tilde{\mathbf{x}})}d\tilde{\mathbf{x}}\right)(1+O(\epsilon_3))\text,
    \end{aligned}
\end{equation}
where $\tilde {U}_\text{min}(\mathbf{x}) = \tilde{U}_0(\mathbf{x}) - \tfrac12\langle\tilde{\mathbf{u}}_1(\mathbf{x}),\tilde{M}^{-1}\tilde{\mathbf{u}}_1(\mathbf{x})\rangle$
is the minimum of $\tilde{U}(\mathbf{x},\mathbf{y})$ at fixed $\mathbf{x}$,
$Q(\tilde{\mathbf{x}}) = \tilde{U}_\text{min}(d\tilde{\mathbf{x}})$ is its geometric part, $d_X$ and $d_Y$ are the dimensions of $X$ and $Y$,
and the error terms are
$\epsilon_1=\tfrac{\beta\mathcal E}{\eps^3d^3}(\|\mathbf{x}\|^2+d\|\mathbf{y}\|)^3+\tfrac{\beta\mathcal E}{d^2\eps^2}\|x\|^3\|y\|+\tfrac{1}{d}(\|\mathbf{x}\|+\|\mathbf{y}\|)$,
and $\epsilon_2$ is given by $\epsilon_1$ by substituting $(a\beta)^{-1/2}$ for $\|\mathbf{y}\|$,
and $\epsilon_3$ by substituting $(a\beta/d^2)^{-1/4}$ for $\|\mathbf{x}\|$.
The final error term is $\epsilon_3 = \beta^{-1/4}\mathcal{E}^{-1/4}d^{-1/2}\eps^{1/2}+\beta^{-1/2}\mathcal{E}^{-1/2}$.

Neglecting the error, this can be written as
\begin{equation}\label{eq:SIZfinal}
    \begin{aligned}
	Z &= Vd^{3N-3} \gamma^{B} \alpha^{-2(3N-6)+d_x} z\text,\\
	z &= (I/d^3)\sigma^{-1}(\det \tilde{M}|_Y)^{-1/2} (2\pi)^{(3N-6-d_X)/2} \int_{X}e^{-Q(\tilde{\mathbf{x}})}d\tilde{\mathbf{x}}\text,
    \end{aligned}
\end{equation}
where $\alpha = (a\beta d^2)^{1/4}$ and $\gamma = e^{\beta \mathcal E}$.

\subsection{Relation to a square-well potential}\label{sec:squarewell}

The calculation presented in the text assumes that the pair potential $V(r)$
has a nonzero second derivative at the minimum. However, many models of sticky spheres,
including Baxter's original calculation, use a square-well pair potential given by
\begin{equation}
    V_\text{square}(r) = \left\{\begin{array}{ll} +\infty & r<d-\epsilon\\ -\mathcal E & d-\epsilon \le r < d+\epsilon \\ 0 & r \ge d+\epsilon \end{array}\right.\text.
\end{equation}
Here we show that if the partition function of a rigid cluster for an analytic potential is given
by $Z_\text{analytic} \sim \exp(-\beta U_\text{min}) \alpha^{-M}$, where $\alpha = (a\beta d^2)^{1/4}$, $a=V''(r)$,
and $M$ is a real exponent, then the partition function for the same cluster
under a square-well pair potential is given by $Z_\text{square} \sim \exp(-\beta U_\text{min}) \alpha^{-M}$, where
$\alpha = (d/\epsilon)^{1/2}$, and $M$ is the same exponent as for $Z_\text{analytic}$. Explicitly,
the notation $f \sim g$ above is used to mean $a g < f < b g$ asymptotically for some constants $a$ and $b$.

Let us show a more general result first. Let $f_1,\ldots, f_B\colon \mathbb{R}^N \to \mathbb{R}$
be real continuous functions such that $f_i(0) = 0$. Define the following integrals:
\begin{equation}
    \begin{aligned}
	Z_1 &=  \int_{\mathbb{R}^N} \prod_{i=1}^{B} \exp(-\tfrac 12 \beta a f_i(\mathbf{r})^2) d\mathbf{r} \text,\\
	Z_2 &=  \int_{\mathbb{R}^N} \prod_{i=1}^{B} \mathbf{1}(-\epsilon\le f_i(\mathbf{r})\le \epsilon) d\mathbf{r} \text,\\
	D_1(t) &= \int_{\mathbb{R}^N} \delta(t - \sum_{i=1}^{B} f_i(\mathbf{r})^2) d\mathbf{r} \text,\\
	D_2(t) &= \int_{\mathbb{R}^N} \delta(t - \max_{i=1,\ldots,B} f_i(\mathbf{r})^2) d\mathbf{r} \text.
    \end{aligned}
\end{equation}
We have 
\begin{equation}
    \begin{aligned}
	Z_1 &= \int_{t=0}^{\infty} D_1(t) \exp(-\tfrac12 \beta a t) dt\\
        Z_2 &= \int_{t=0}^{\epsilon^{2}} D_2(t) dt\text.
    \end{aligned}
\end{equation}
We also have that
\begin{equation}
    \int_{t'=0}^{t/B} D_2(t') dt' \le \int_{t'=0}^{t} D_1(t') dt' \le \int_{t'=0}^{t} D_2(t') dt'\text,
\end{equation}
since $\max f_i^2 \le\sum f_i^2 \le B\max f_i^2$. Therefore, we have the following sequence of implications:
\begin{equation}
    Z_1 \sim (\beta a)^{-M} \;\;\Leftrightarrow\;\; D_1(t) \sim t^{M-1} \;\;\Leftrightarrow\;\; D_2(t) \sim t^{M-1} \;\;\Leftrightarrow\;\; Z_2 \sim \epsilon^{2M}\text.
\end{equation}

As a special case, we have that if $$Z_\text{analytic} = V d^{3N-3} (2\pi)^{-B_\text{ISO}/2} \exp(B\beta\mathcal E) (a\beta d^2)^{-M} z\text,$$
then $$Z_\text{square} = V d^{3N-3} (2\pi)^{-B_\text{ISO}/2} \exp(B\beta\mathcal E) (\epsilon/d)^{2M} z'\text.$$ However, the geometric
parts $z$ and $z'$ can be different from each other for the two potentials.

\end{widetext}

\section{Calculating the partition functions}

\subsection{Numerical method to calculate the geometric partition functions}\label{sec:numerical}

Evaluating most of the quantities in equation \eqref{eq:zg} is straightforward. The two quantities that are worth discussing are the integral over the singular subspace, and the symmetry number. 

\subsubsection{Integrating the exponential of a quartic}

Here is how we numerically calculate
\begin{equation}\label{eq:Q2}
\int_X e^{-Q(\tilde{\mathbf{x})}}d\tilde{\mathbf{x}}.
\end{equation}

If $d_X = 1$, then we use the fact that $\int_{-\infty}^\infty e^{-x^4}dx = 2\Gamma(5/4)$ to write \eqref{eq:Q2} as $2\Gamma(5/4)(Q(\mathbf{v}))^{-1/4}$, where $\mathbf v \in X$ is a unit vector.

If $d_X \geq 2$ we integrate \eqref{eq:Q2} numerically. While one can derive analytic expressions for the integral, even for a two-dimensional integral these are unwieldy \cite[][see e.g. Table 2, formula for $n=2,r=3$]{Morozov2009}. 
Our numerical method is as follows: form an orthogonal basis $\{\mathbf{v}_i\}_{i=1}^{d_X}$ of $X$. For each direction $\mathbf{v}_i$, determine a value $s_i$ such that $Q(s_i\mathbf{v}_i) \in [10^{-19},10^{-16}]$. 
Then for $2\leq d_X\leq 4$ we integrate over a box $[-s_1,s_1]\times \ldots \times[-s_{d_X}, s_{d_X}]$ using the trapezoidal rule  with equally-spaced points. For $d_X=5$ we use a Monte-Carlo method where we choose points either uniformly, or do importance sampling from a Gaussian with standard deviation $s_i/3$ in each direction. 

We use 25 points per dimension for the deterministic integrals, which gives us an error between $10^{-6}$ and $10^{-7}$. We test this using Matlab's built-in function \texttt{integral}, which has a default accuracy of amount that amount. 
For the single integral with $d_X=5$ we use $10^6$ points, so the error is expected to be about $10^{-3}$. 
Using the trapezoidal rule we obtain excellent accuracy with very few points. For smooth periodic functions the trapezoidal rule achieves exponential convergence once enough discretization nodes are used so as to sample at the Nyquist rate. For non-periodic functions, the error is dominated by the derivatives of the integrand at its endpoints, and achieves at least second-order convergence \cite{Trefethen:2014du}.

\subsubsection{Symmetry numbers} 

The symmetry number accounts for the total number of distinct copies of each cluster obtained by permuting the labeling of the spheres.
In our case we count two clusters as ``distinct'' if there is no rotation that maps the labeled spheres of one cluster to the spheres with
the same labels in the other cluster.

The total number of copies of a rigid cluster of $N$ spheres (including enantiomers) is
\begin{equation}
 \frac{2N! }{a_k},
\end{equation}
where $a_k$ is the total number of permutations that map an adjacency matrix to itself and also preserve all the pairwise distances.
The symmetry number for this method of counting is $\sigma=a_k$.
We calculate $a_k$ by first computing the automorphism group of the adjacency matrix using the function \texttt{allgroup3} in the program \texttt{nauty} \cite{mckay1981}. 
For each element in the automorphism group, we apply the corresponding permutation to the particles and check if the pairwise distance matrix is preserved. If so, we increase $a_k$ by 1. 
We have to check the pairwise distances, because a permutation that preserves the adjacency matrix can yield a cluster that is distinct from the original. For example,
consider a large octahedral shell, made of triangles glued together. Pick one triangle, and attach two spheres to it, one above and one below the triangle, to form a bipyramid. There is an automorphism of the adjacency matrix, namely switching the two spheres, that does not correspond to a rotation or a reflection.


\subsection{Simulation method}\label{sec:sims}

We performed Metropolis Monte Carlo simulations for $N=9$ spheres interacting via a Morse-Harmonic potential with range parameter $\rho$ of the form
\begin{equation}\label{eq:morse}
    V(r) = \left\{\begin{array}{ll} \rho^2 (r-1)^2 - 1 & r\le 1 \\ \exp(-\rho(r-1))[\exp(-\rho(r-1))-2] & r>1\end{array}\right.
\end{equation}
We replaced the Morse potential with a spring potential for $r<1$ for numerical stability reasons. The entire matched potential is continuous with continuous first and second derivatives. 
We use a periodic simulation cell of size $6\times6\times6$.

Let $\mathbf{r}(t)$ be the coordinates sampled from the Monte Carlo trajectory at time $t$.
We construct an adjacency matrix such that $a_{ij}(t) = 1$ if $r_{ij}<r_\text{cutoff}$ and $a_{ij}=0$ otherwise.
We set $b_k(t) = 1$ for the cluster $1\le k\le52$ (if any) whose adjacency matrix $A_k$ is isomorphic to $A(t)$
and $b_k(t)=0$ for all others. The frequency of cluster $k$ is $P_k = \langle b_k\rangle / \sum_{k'=1}^{52} \langle b_{k'}\rangle$.
We also measure the correlation function $C_k(\tau) = \langle b_k(t)b_k(t+\tau)\rangle / \langle b_k\rangle^2$.
The estimated error for $P_k$ is $\Delta P_k/P_k = (\langle b_k\rangle T / \tau_k)^{-1/2}$, where
$T$ is the length of the simulation and $\tau_k$ is the correlation time, which we estimate as
$\tau_k = \sum_{\tau=0}^\infty C_k(\tau)$.

\begin{figure*}
    \begin{center}
	\begin{tikzpicture}
	    \begin{axis}[
		    width=0.45\linewidth,scaled ticks=false,
		    xlabel={$\alpha$}, ylabel={$P_{11}$},
		    xmin=11, xmax=26,
		    ymin=0, ymax=0.06,
		    domain=11:26,
		    legend style={at={(0.97,0.03)},anchor=south east},
		    legend entries={cut off,$1+1/\rho$,$1+3/\rho$,$1+5/\rho$},
		    yticklabel style = { /pgf/number format/fixed },
		]
		\addlegendimage{empty legend}
	        \addplot[only marks, mark size=1.0pt, error bars/y dir=both,error bars/y explicit,color=orange] table[x index=0, y index=1, y error index =2] {n9-k200-i11.dat};
	        \addplot[only marks, mark size=1.0pt, error bars/y dir=both,error bars/y explicit,color=red] table[x index=0, y index=3, y error index =4] {n9-k200-i11.dat};
	        \addplot[only marks, mark size=1.0pt, error bars/y dir=both,error bars/y explicit,color=purple] table[x index=0, y index=5, y error index =6] {n9-k200-i11.dat};
		\addplot[dotted,color=black]plot (\x,{8.85016/(235.4+\x)});
	    \end{axis}
	\end{tikzpicture}
	\begin{tikzpicture}
	    \begin{axis}[
		    width=0.45\linewidth,scaled ticks=false,
		    xlabel={$\alpha$}, ylabel={$P_{19}$},
		    xmin=11, xmax=26,
		    ymin=0, ymax=0.06,
		    domain=11:26,
		    legend style={at={(0.97,0.03)},anchor=south east},
		    legend entries={cut off,$1+1/\rho$,$1+3/\rho$,$1+5/\rho$},
		    yticklabel style = { /pgf/number format/fixed },
		]
		\addlegendimage{empty legend}
	        \addplot[only marks, mark size=1.0pt, error bars/y dir=both,error bars/y explicit,color=orange] table[x index=0, y index=1, y error index =2] {n9-k200-i19.dat};
	        \addplot[only marks, mark size=1.0pt, error bars/y dir=both,error bars/y explicit,color=red] table[x index=0, y index=3, y error index =4] {n9-k200-i19.dat};
	        \addplot[only marks, mark size=1.0pt, error bars/y dir=both,error bars/y explicit,color=purple] table[x index=0, y index=5, y error index =6] {n9-k200-i19.dat};
		\addplot[dotted,color=black]plot (\x,{8.53408/(235.4+\x)});
	    \end{axis}
	\end{tikzpicture}\\
	\begin{tikzpicture}
	    \begin{axis}[
		    width=0.45\linewidth,scaled ticks=false,
		    xlabel={$\alpha$}, ylabel={$P_{28}$},
		    xmin=11, xmax=26,
		    ymin=0, ymax=0.06,
		    domain=11:26,
		    legend style={at={(0.97,0.03)},anchor=south east},
		    legend entries={cut off,$1+1/\rho$,$1+3/\rho$,$1+5/\rho$},
		    yticklabel style = { /pgf/number format/fixed },
		]
		\addlegendimage{empty legend}
	        \addplot[only marks, mark size=1.0pt, error bars/y dir=both,error bars/y explicit,color=orange] table[x index=0, y index=1, y error index =2] {n9-k200-i28.dat};
	        \addplot[only marks, mark size=1.0pt, error bars/y dir=both,error bars/y explicit,color=red] table[x index=0, y index=3, y error index =4] {n9-k200-i28.dat};
	        \addplot[only marks, mark size=1.0pt, error bars/y dir=both,error bars/y explicit,color=purple] table[x index=0, y index=5, y error index =6] {n9-k200-i28.dat};
		\addplot[dotted,color=black]plot (\x,{8.85016/(235.4+\x)});
	    \end{axis}
	\end{tikzpicture}
	\begin{tikzpicture}
	    \begin{axis}[
		    width=0.45\linewidth,scaled ticks=false,
		    xlabel={$\alpha$}, ylabel={$P_{42}$},
		    xmin=11, xmax=26,
		    ymin=0, ymax=0.06,
		    domain=11:26,
		    legend style={at={(0.97,0.97)},anchor=north east},
		    legend entries={cut off,$1+1/\rho$,$1+3/\rho$,$1+5/\rho$},
		    yticklabel style = { /pgf/number format/fixed },
		]
		\addlegendimage{empty legend}
	        \addplot[only marks, mark size=1.0pt, error bars/y dir=both,error bars/y explicit,color=orange] table[x index=0, y index=1, y error index =2] {n9-k200-i42.dat};
	        \addplot[only marks, mark size=1.0pt, error bars/y dir=both,error bars/y explicit,color=red] table[x index=0, y index=3, y error index =4] {n9-k200-i42.dat};
	        \addplot[only marks, mark size=1.0pt, error bars/y dir=both,error bars/y explicit,color=purple] table[x index=0, y index=5, y error index =6] {n9-k200-i42.dat};
		\addplot[dotted,color=black]plot (\x,{4.42508/(235.4+\x)});
	    \end{axis}
	\end{tikzpicture}
\caption{Simulation results for $N=9$ sticky spheres with $\kappa\approx220$ and $\rho=30,40,50,60,70,80,100,120,$ and $140$ interpreted
    using different bond cut offs. We show the observed frequencies of four of the 51 nonsingular clusters. The dashed line shows the theoretically predicted frequency. 
    Compare to  Figure \ref{fig:sims}, which plots the same data for the singular cluster. \label{fig:SIsims}}
\end{center}
\end{figure*}
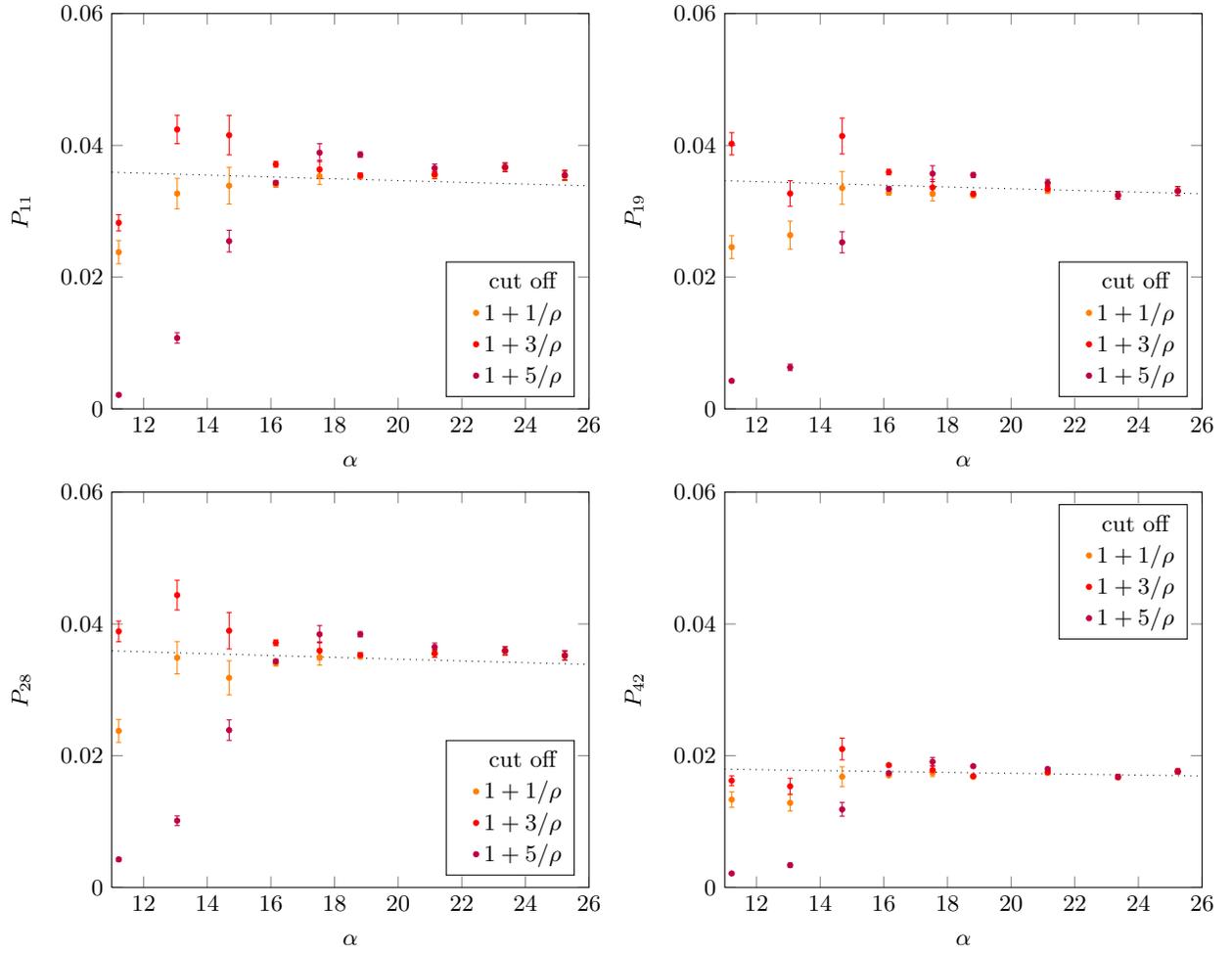

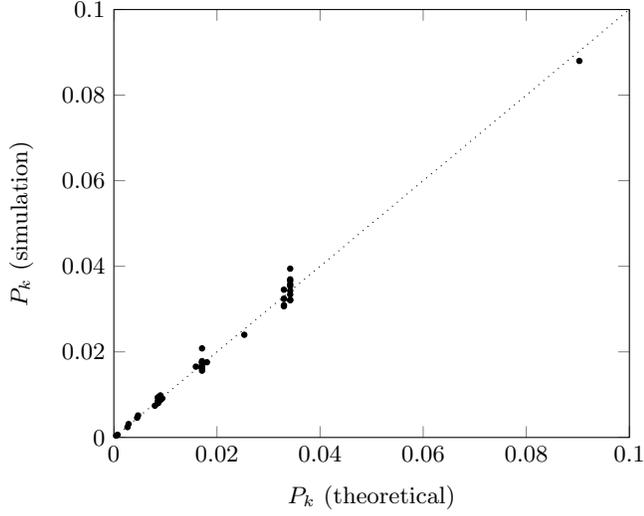
\begin{figure}
    \begin{center}
	\scalebox{1.0}{\begin{tikzpicture}
	    \begin{axis}[
		    xlabel={$P_k$ (theoretical)}, ylabel={$P_k$ (simulation)},
		    xmin=0.0, xmax=0.1,
		    ymin=0.0, ymax=0.1,
		    domain = 0.0:0.1,
		    yticklabel style = { /pgf/number format/fixed },
		    xticklabel style = { /pgf/number format/fixed },
		]
	        \addplot[only marks, mark size=1.0pt] table[x index=0, y index=1] {p120-scatter.dat};
		\addplot[dotted,color=black]plot (\x,{\x});
	    \end{axis}
	\end{tikzpicture}}
\caption{Predicted vs.\ observed frequencies for the 52 different rigid cluster geometries for $N=9$ sticky spheres with $\kappa\approx220$ and $\rho=120$,
    using a cut off of $1+1/\rho$. \label{fig:scatter}}
\end{center}
\end{figure}

In the main text, we focus on the observed and predicted frequencies of the singular cluster. For completeness,
we show in  Figure \ref{fig:SIsims} and  Figure \ref{fig:scatter} some of the results for the nonsingular clusters. 
 Figure \ref{fig:SIsims} shows that the difficulty of identifying a cluster for larger values of the range parameter, occurs just as much for regular clusters as for singular ones. 

\subsection{Scaling Laws}\label{sec:scalings}

\begin{figure*}
\begin{center}
\includegraphics[trim={1cm 0 4cm 0},clip,width=0.9\linewidth]{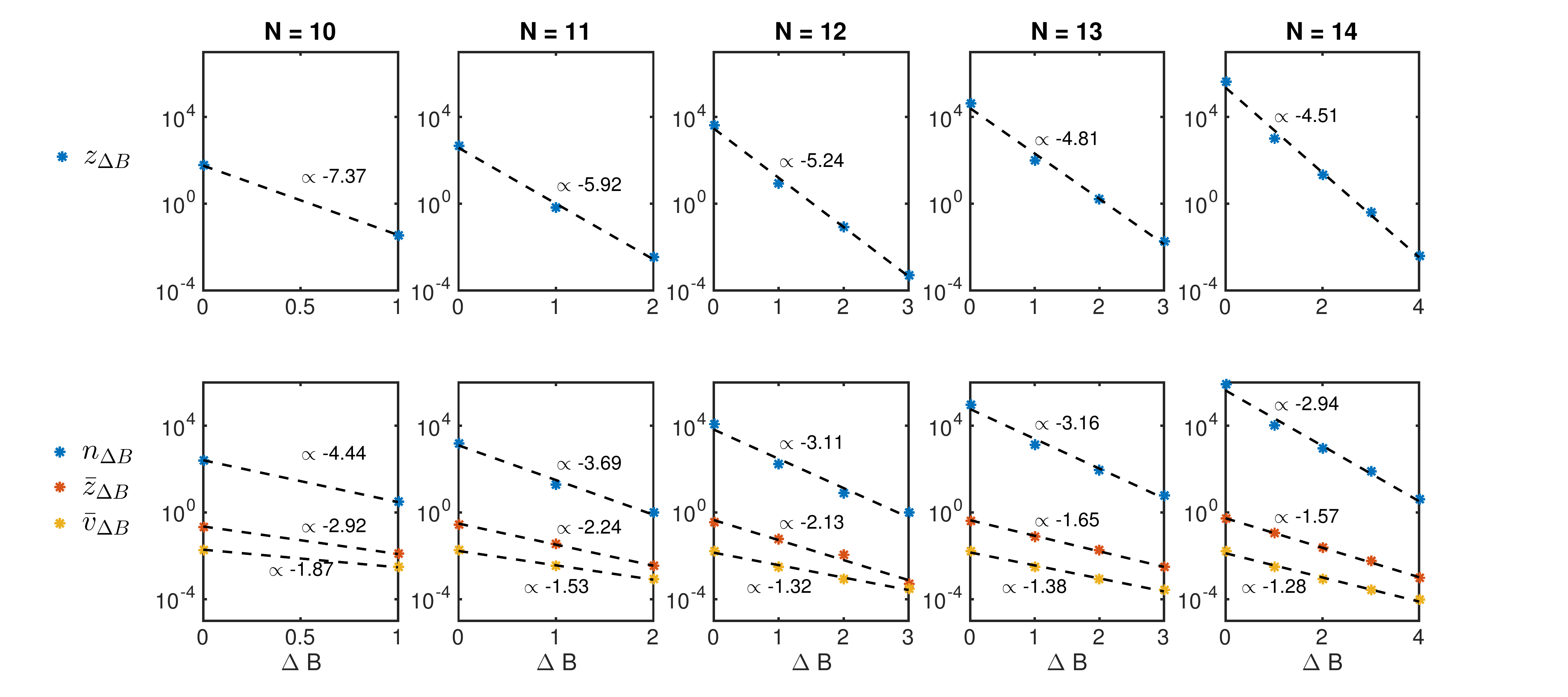}
\end{center}
\caption{Scaling laws for $N=10-14$. Same labeling as  Figure \ref{fig:scalings}.\label{fig:scalings2}}
\end{figure*}

 Figure \ref{fig:scalings2} shows the same scaling laws as  Figure \ref{fig:scalings}, but for smaller $N$. 

\begin{figure*}
\begin{center}
\includegraphics[trim={1cm 0 4cm 0},clip,width=0.9\linewidth]{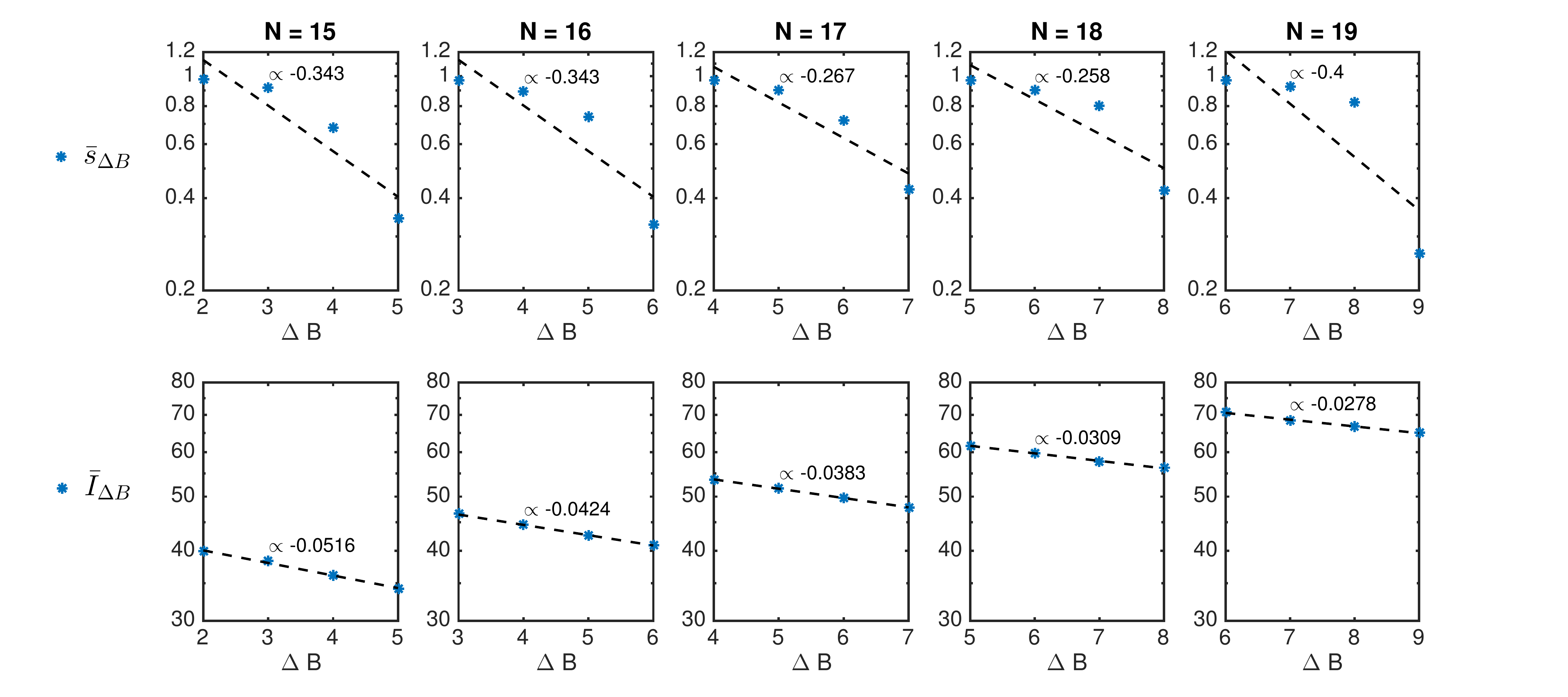}\\
\includegraphics[trim={1cm 0 4cm 0},clip,width=0.9\linewidth]{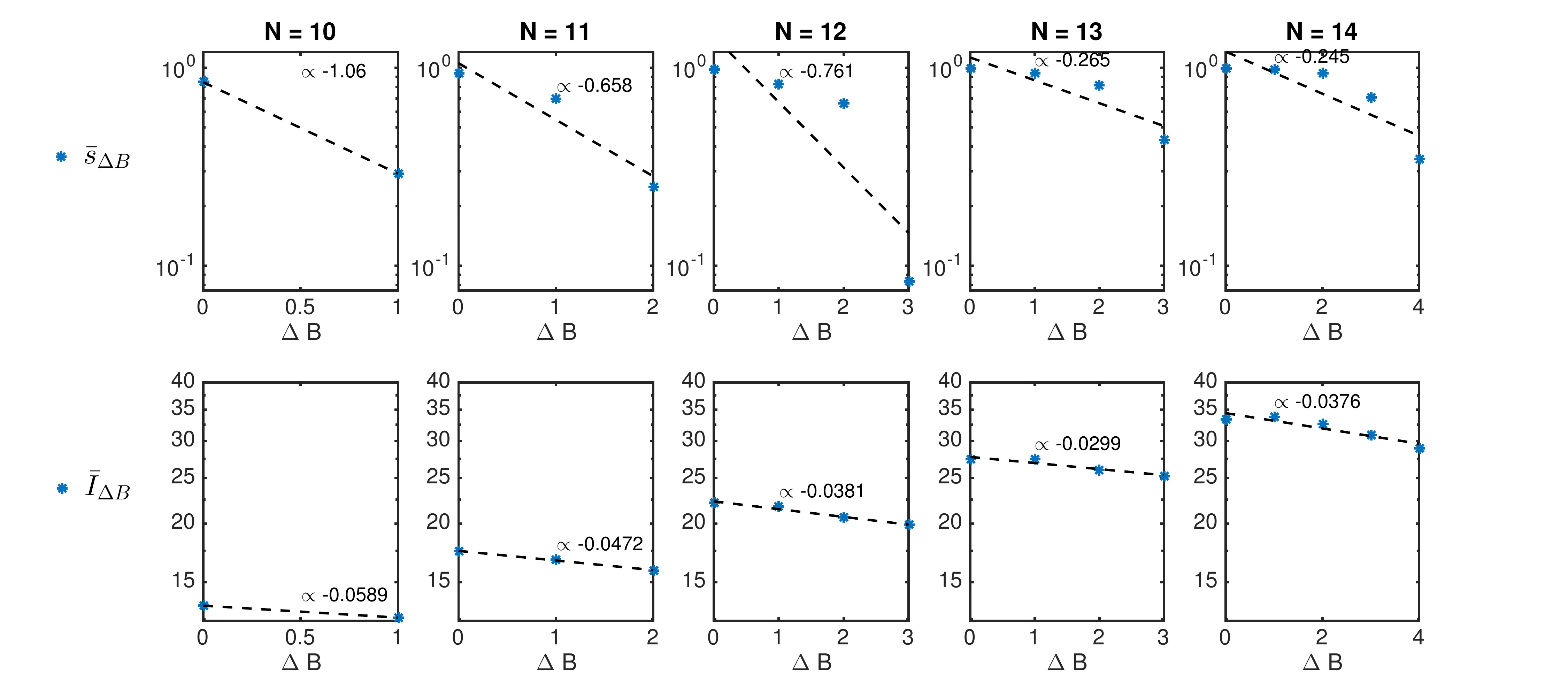}
\end{center}
\caption{Mean symmetry number $1/\sigma$ and mean moment of inertia factor $|I|^{1/2}$ for regular clusters as a function of $\Delta B$, for each $N=10-19$. The dashed line is the best-fit line with best-fit slope shown beside it. Vertical axis is a logarithmic scale. This shows that neither factor contributes strongly to the exponential scaling of $z_{\Delta B}$.}\label{fig:scalings3}
\end{figure*}

 Figure \ref{fig:scalings3} shows the mean inverse symmetry number $\bar{s} _{\Delta B} = \oneover{n_{\Delta B}}\sum_{}\sigma^{-1}$ for regular clusters with a given $\Delta B$, and the mean moment of inertia factor $\bar{I}_{\Delta B} = \oneover{n_{\Delta B}}\sum_{}I$. The mean inverse symmetry number does not follow an exponential scaling law. The mean moment of inertia does, but the slope is tiny. This demonstrates that neither the symmetry number, nor the moment of inertia, are important factors in determining why $\bar{z}_{\Delta B}$ follows an exponential scaling. 

\subsection{Bond-orientational order parameter}\label{sec:q6}
Given a cluster $\mathbf{r}$, with bonds $E=\{(i_1,j_1),\ldots,(i_B,j_B)\}$,
we define the bond-orientational order parameter $Q_l$ as follows:
\begin{equation}
    \begin{aligned}
	q_{l,m} &= \frac{1}{B} \sum_{(i,j)\in E} Y_{l,m}(\mathbf{r}_{ij}/d)\\
	Q_l &= \left(\frac{1}{2l+1} \sum_{m=-l}^l q_{l,m}^2\right)^{1/2}\text,
    \end{aligned}
\end{equation}
where $Y_{l,m}(\mathbf{u})$, $m=-l,\ldots,l$, is a basis for the spherical harmonics of
degree $l$ normalized such that $\int_{S^2} Y_{l,m}(\mathbf{u})^2d\mathbf{u} = 4\pi$.

The value of the bond-orientational order parameter $Q_6$ for a large fragment of the face-centered
cubic lattice approaches the value $0.575$ (the limit is $0.485$ for the hexagonal close-packed lattice).
On the other hand, if the bond directions are drawn randomly from a uniform measure on the sphere,
the order parameter will tend to zero.

\begin{widetext}

\clearpage

\section{Values of the geometric partition functions}\label{sec:values}

The following tables give the values of $z_{\Delta B, d_X}$ for each value of $N$. Each value is shown to 3 significant digits (if fewer are shown, it is because the trailing significant digits are zero.) 

\paragraph{$n=6$}
\begin{tabular}{c||c}
$d_X$ $\backslash$ $\Delta B$ & 0 \\\hline\hline
0  & 0.0501 \\\hline
\end{tabular}
\bigskip

\paragraph{$n=7$}
\begin{tabular}{c||c}
$d_X$ $\backslash$ $\Delta B$ & 0 \\\hline\hline
0  &  0.222 \\\hline
\end{tabular}
\bigskip

\paragraph{$n=8$}
\begin{tabular}{c||c}
$d_X$ $\backslash$ $\Delta B$ & 0 \\\hline\hline
0  &   1.22 \\\hline
\end{tabular}
\bigskip

\paragraph{$n=9$}
\begin{tabular}{c||c}
$d_X$ $\backslash$ $\Delta B$ & 0 \\\hline\hline
0  &   7.88 \\\hline
1  & 0.0335 \\\hline
\end{tabular}
\bigskip

\paragraph{$n=10$}
\begin{tabular}{c||c|c|c}
$d_X$ $\backslash$ $\Delta B$ & -1  & 0  & 1 \\\hline\hline
0  & .  &   57.1  & 0.0362 \\\hline
1  & .  &  0.427  & . \\\hline
2  &   0.27  & .  & . \\\hline
\end{tabular}
\bigskip

\paragraph{$n=11$}
\begin{tabular}{c||c|c|c|c|c}
$d_X$ $\backslash$ $\Delta B$ & -2  & -1  & 0  & 1  & 2 \\\hline\hline
0  & .  & .  &    456  &  0.701  & 0.00328 \\\hline
1  & .  & .  &   4.05  & 0.0111  & . \\\hline
2  & .  &   5.37  & .  & .  & . \\\hline
3  &  0.771  & .  & .  & .  & . \\\hline
\end{tabular}
\bigskip

\paragraph{$n=12$}
\begin{tabular}{c||c|c|c|c|c|c}
$d_X$ $\backslash$ $\Delta B$ & -2  & -1  & 0  & 1  & 2  & 3 \\\hline\hline
0  & .  & .  & 4.06e+03  &   9.24  & 0.0898  & 0.000501 \\\hline
1  & .  & .  &   35.8  &  0.225  & .  & . \\\hline
2  & .  &   59.5  &  0.158  & .  & .  & . \\\hline
3  &    9.5  & .  & .  & .  & .  & . \\\hline
\end{tabular}
\bigskip

\paragraph{$n=13$}
\begin{tabular}{c||c|c|c|c|c|c}
$d_X$ $\backslash$ $\Delta B$ & -2  & -1  & 0  & 1  & 2  & 3 \\\hline\hline
0  & .  & .  & 4.06e+04  &    101  &   1.65  & 0.0176 \\\hline
1  & .  & .  &    345  &    2.7  & 0.00517  & 0.000482 \\\hline
2  & .  &    604  &   2.07  & .  & .  & . \\\hline
3  &   90.9  &    0.5  & .  & .  & .  & . \\\hline
4  &  0.221  & .  & .  & .  & .  & . \\\hline
\end{tabular}
\bigskip

\paragraph{$n=14$}
\begin{tabular}{c||c|c|c|c|c|c|c|c}
$d_X$ $\backslash$ $\Delta B$ & -3  & -2  & -1  & 0  & 1  & 2  & 3  & 4 \\\hline\hline
0  & .  & .  & .  & 4.56e+05  & 1.04e+03  &     22  &  0.417  & 0.0037 \\\hline
1  & .  & .  & .  & 3.71e+03  &   29.8  & 0.0791  & 0.00488  & . \\\hline
2  & .  & .  & 6.41e+03  &   21.7  & .  & .  & .  & . \\\hline
3  & .  &    918  &   11.7  & .  & .  & .  & .  & . \\\hline
4  & .  &   8.25  & .  & .  & .  & .  & .  & . \\\hline
5  &  0.968  & .  & .  & .  & .  & .  & .  & . \\\hline
\end{tabular}
\bigskip

\paragraph{$n=15$}
\begin{tabular}{c||c|c|c|c}
$d_X$ $\backslash$ $\Delta B$ & 2  & 3  & 4  & 5 \\\hline\hline
0  &    250  &   6.64  &  0.112  & 0.00163 \\\hline
1  &   1.17  & 0.0471  & .  & . \\\hline
\end{tabular}
\bigskip

\paragraph{$n=16$}
\begin{tabular}{c||c|c|c|c}
$d_X$ $\backslash$ $\Delta B$ & 3  & 4  & 5  & 6 \\\hline\hline
0  &   87.8  &   2.14  & 0.0458  & 0.000727 \\\hline
1  &  0.611  & .  & .  & . \\\hline
\end{tabular}
\bigskip

\paragraph{$n=17$}
\begin{tabular}{c||c|c|c|c}
$d_X$ $\backslash$ $\Delta B$ & 4  & 5  & 6  & 7 \\\hline\hline
0  &     33  &  0.898  & 0.0223  & 0.000257 \\\hline
1  & 0.0304  & .  & .  & . \\\hline
\end{tabular}
\bigskip

\paragraph{$n=18$}
\begin{tabular}{c||c|c|c|c}
$d_X$ $\backslash$ $\Delta B$ & 5  & 6  & 7  & 8 \\\hline\hline
0  &   14.7  &  0.462  & 0.00989  & 8.38e-05 \\\hline
1  &  0.012  & .  & .  & . \\\hline
\end{tabular}
\bigskip

\paragraph{$n=19$}
\begin{tabular}{c||c|c|c|c}
$d_X$ $\backslash$ $\Delta B$ & 6  & 7  & 8  & 9 \\\hline\hline
0  &    7.8  &   0.24  & 0.00408  & 3.39e-05 \\\hline
1  & 0.0107  & .  & .  & . \\\hline
\end{tabular}
\bigskip

\end{widetext}

\end{document}